\documentclass[paper,onecolumn,twoside]{geophysics}

\date{} 

\usepackage{enumitem}

\usepackage{url}

\usepackage{hyperref} 

\renewcommand{\figdir}{Fig}
\newcommand*{\Scale}[2][4]{\scalebox{#1}{$#2$}} 

\usepackage{amssymb}
\usepackage{mathrsfs}
\usepackage{amsmath}
\usepackage{algorithm,algorithmic}
\usepackage{url}
\usepackage{multirow}
\usepackage{threeparttable}
\makeatletter
\newcommand{\ALOOP}[1]{\ALC@it\algorithmicloop\ #1%
  \begin{ALC@loop}}
\newcommand{\ENDALOOP}{\end{ALC@loop}\ALC@it\algorithmicendloop}

\usepackage{setspace}
\usepackage[skip=2pt, font=normalsize]{caption}
\makeatother

\begin{document}

\onecolumn 

\begin{description}[labelindent=0cm,leftmargin=4cm,style=multiline]

\item[\textbf{Citation}]{Z. Wang, T. Hegazy, Z. Long and G. AlRegib, ``Noise-robust detection and tracking of salt domes in postmigrated volumes using texture, tensors, and subspace learning,'' Geophysics, vol. 80, no. 6, pp. WD101-WD116, Sep. 2015.}

\item[\textbf{DOI}]{\url{https://doi.org/10.1190/geo2015-0116.1}}

\item[\textbf{Review}]{Date of publication: 23 September 2015}

\item[\textbf{Bib}] {@article\{wang2015noise,\\
  title=\{Noise-robust detection and tracking of salt domes in postmigrated volumes using texture, tensors, and subspace learning\},\\
  author=\{Wang, Z. and Hegazy, T. and Long, Z. and AlRegib, G.\},\\
  journal=\{Geophysics\},\\
  volume=\{80\},\\
  number=\{6\},\\
  pages=\{WD101--WD116\},\\
  year=\{2015\},\\
  publisher=\{Society of Exploration Geophysicists\}\}
} 


\item[\textbf{Copyright}]{\textcopyright 2015 SEG. Personal use of this material is permitted. Permission from SEG must be obtained for all other uses, in any current or future media, including reprinting/republishing this material for advertising or promotional purposes, creating new collective works, for resale or redistribution to servers or lists, or reuse of any copyrighted component of this work in other works.}

\item[\textbf{Contact}]{\href{mailto:zhiling.long@gatech.edu}{zhiling.long@gatech.edu}  OR \href{mailto:alregib@gatech.edu}{alregib@gatech.edu}\\ \url{https://ghassanalregib.com/} \\ }
\end{description}

\thispagestyle{empty}
\newpage
\clearpage
\setcounter{page}{1}


\title{Noise-robust detection and tracking of salt domes in postmigrated volumes using texture, tensors, and subspace learning}

\renewcommand{\thefootnote}{\fnsymbol{footnote}}

\address{
\footnotemark[1] Center for Energy and Geo Processing (CeGP) at Georgia Tech and KFUPM,
School of Electrical and Computer Engineering,
Georgia Institute of Technology, Atlanta, GA 30332-0250}
\author{Zhen Wang\footnotemark[1], Tamir Hegazy\footnotemark[1], Zhiling Long\footnotemark[1], and Ghassan AlRegib\footnotemark[1]}

\lefthead{Wang et al.}
\righthead{Salt Dome Detection and Tracking}

\begin{abstract}
The identification of salt dome boundaries in migrated seismic data volumes is important for locating petroleum reservoirs. The presence of noise in the data makes computer-aided salt dome interpretation even more challenging. In this paper, we develop noise-robust algorithms that can label boundaries of salt domes both effectively and efficiently. Our research is twofold. First, we utilize a texture-based gradient to accomplish salt dome detection. We show that by employing a dissimilarity measure based on two-dimensional (2D) discrete Fourier transform (DFT), the algorithm is capable of efficiently detecting salt dome boundaries with accuracy. At the same time, our analysis shows that the proposed algorithm is robust to noise. Once the detection is performed for an initial 2D seismic section, we propose to track the initial boundaries through the data volume to accomplish an efficient labeling process by avoiding parameters tuning that would have been necessary if detection had been performed for every seismic section. The tracking process involves a tensor-based subspace learning process, in which we build texture tensors using patches from different seismic sections. To accommodate noise components with various levels in a texture tensor, we employ noise-adjusted principal component analysis (NA-PCA), so that principal components corresponding to greater signal-to-noise ratio values may be selected for tracking. We validate our detection and tracking algorithms through experiments using seismic datasets acquired from Netherland offshore F3 block in the North Sea with very encouraging results. 
\end{abstract}

\section{Introduction}
\label{sec:intro}
%
%
%
The deposition of salt in marine basins commonly intrudes into surrounding rock strata and forms an important geological structure, salt domes. Because of the impermeability of salt, salt domes may seal porous reservoir rocks and lead to the formation of petroleum reservoirs. To localize the possible positions of reservoir regions, experienced interpreters need to accurately label the boundaries of salt domes in migrated seismic data. With the dramatically increasing amount of acquired seismic data, however, manual interpretation is becoming time consuming and labor intensive.

To speed up interpretation, in recent years, researchers have proposed various interactive methods that assist with the detection of salt dome boundaries using graph theory and image processing techniques. \cite{lomask2004image} defined seismic sections as weighted undirected graphs and applied the normalized cut image segmentation (NCIS)~\cite[]{shi2000normalized} to globally optimize the delineation of salt dome boundaries. As an extension of this method, \cite{lomask2007application} employed local dips in the weight matrix and utilized bound constraints to remove boundary artifacts. Similarly, \cite{halpert2009seismic} introduced the modified NCIS by combining multiple seismic attributes with adaptive weights. Although these NCIS-based methods can be implemented in parallel, the high computational cost limits their future application on high-resolution or three-dimensional (3D) seismic data. To improve the efficiency of global segmentation, \cite{halpert2010speeding} proposed to detect the boundaries of salt domes using a pairwise region comparison based on the minimum spanning tree~\cite[]{felzenszwalb2004efficient}, which reduces the algorithm complexity from $\mathcal{O}\left(n^2\right)$ to $\mathcal{O}\left(n \log n\right)$.

In addition to the segmentation-based methods, edge detection has also been explored. \cite{jing2007detecting} and \cite{aqrawi2011detecting} applied dip-guided 2D and 3D Sobel filters to detect salt dome boundaries in time slices. In addition, \cite{berthelot2013texture} proposed to delineate salt domes by testing a combination of multiple attributes (i.e., texture, frequency, and dip) in a supervised Bayesian classification model. To characterize the homogeneous texture of salt bodies, most recently, \cite{hegazy2014texture} derived a directionality attribute from the moment of inertia tensor defined by gradient components. Although the methods mentioned above may be able to achieve various levels of improvement in the detection performance, normally they involve many parameters to be manually tuned, which unavoidably compromises the overall efficiency. In addition to the effectiveness and efficiency issues, salt dome interpretation may be further complicated by the presence of noise. Typically, noise from different sources is present, which may vary across a seismic volume, thus being space-time dependent. Accommodating such noise effect is a challenging task for computer-aided interpretation.

In our work, we view seismic volumes as big data volumes containing certain information that needs to be extracted manually or automatically. Hence, we import our long experience in digital signal processing and visual data analysis to seismic interpretation. In our view, a large seismic volume contains information about salt dome structures. Such structure is characterized by certain attributes. However, calculating these attributes accurately and efficiently faces many challenges. First, the data are subjected to a long processing pipeline of stacking and migration. Not only do these steps introduce artifacts, but they may magnify noise in the raw data. Second, the raw data are an undersampled depiction of the subsurface structure. This is analogous to detecting detailed activities from low-quality videos captured with a low-resolution camera. Third, noise in the data volume can be a function of the subsurface structure, and it may contain information about the subsurface structure. Instead of denoising, our long-term plan is to take advantage of such noise.

In this paper, we focus on developing noise-robust algorithms that can label boundaries of salt domes both effectively and efficiently. Our research is twofold. First, we define a texture-based attribute, namely gradient of texture (GoT), and combine it with a dissimilarity measure based on 2D discrete Fourier transform (DFT) to implement an algorithm capable of detecting salt dome boundaries with high accuracy and efficiency. We further analyze and show that the proposed algorithm is robust to noise. Once salt dome detection is performed for an initial 2D seismic section, we propose to track the initial boundaries through the data volume to accomplish the labeling process. This way, we can avoid tweaking parameters for boundary detection in every seismic section, thus being more efficient. The tracking involves a tensor-based subspace learning process. It starts by forming texture tensors according to the initially labeled boundaries, to better capture the strong correlation between patches of similar texture patterns. It then projects the initial boundaries onto the neighboring seismic section, and determines the tracked boundaries by evaluating texture similarity in the subspace of the texture tensors. To maintain consistency, we rename initial and neighboring seismic sections as reference and predicted sections, respectively, in the rest of this paper. To apply the tensor-based subspace learning, we also take noise into consideration. Unlike traditional applications, with seismic data it is likely that noise of different levels may be included in the same texture tensor. To handle such situations, we employ noise-adjusted subspace learning techniques such as those presented in~\cite[]{roger1994faster, chang1999interference}.

An overall block diagram for our salt dome boundary labeling system is depicted in Figure~\ref{fig:diagram}. It includes both the detection of salt dome boundaries for the reference sections, and the tracking of the boundaries through the predicted sections. This figure will be used throughout the paper. The rest of the paper is organized as follows. First, we discuss our salt dome detection method. Then, we present our salt dome tracking algorithm. Thereafter, we conduct experiments to evaluate the performance of the proposed methods and the noise effect. Based on the results, we draw conclusions at the end of the paper.


\section{Theory}
\subsection{Salt Dome Detection}
\label{sec:detection}

The top boundary of a salt dome usually has a cap rock, which appears in migrated seismic surveys as areas with high contrast and strong edges (Figure~\ref{fig:SaltDome}). However, the lateral boundaries usually lack such high contrast and are more difficult to detect.  Figure~\ref{fig:SaltDome} shows that the lateral boundaries of the salt dome have
%
%
no explicit edge existing between the salt region and the non-salt region containing horizons. However, seismic interpretation experts would still be able to visually delineate such boundaries using the change of texture between salt and non-salt regions shown by other attributes. As a matter of fact, even non-experts can still perceptually differentiate between the two regions and possibly draw a boundary line. Therefore, we believe that an automated method should take human visual perception into account.  Using traditional edge detection methods (e.g., Sobel filter) to delineate such boundary would fail because the boundary exhibits a change in texture, rather than an explicit edge. 

Figure~\ref{fig:diagram} shows an overview of the proposed detection method. To capture the change in texture, we propose a new attribute, which we call \textit{Gradient of Texture} (GoT). Then, a threshold is applied to the GoT map $\mathbf{G}$ to obtain a binary image $\mathbf{B_t}$. From an initialization point inside the salt dome, we perform region growing such that the grown region boundaries stop at the threshold value of GoT.  We perform binary morphological operations on the grown region to  enhance it. The boundary of the salt region is determined based on enhanced grown region. In the following, we explain each of the modules in detail.

In relation to existing methods from the image processing literature, we show in the following subsections that our proposed GoT is faster and more general  than the ``texture gradient'' introduced by~\cite{MF+04}. First, \cite{MF+04} computes features along 13 different directions, while we show that only 2 directions (horizontal and vertical) are sufficient. Second,~\cite{MF+04} computes the gradient as the norm of the difference between 13-dimensional feature vectors while we directly assess the dissimilarity between the neighborhood windows without computing feature vectors. Finally,~\cite{MF+04} compares texture in two disk halves centered around a center point, while we use square neighborhood windows sharing a side centered around the center point. Using square (or even rectangular)  windows makes it easier to take advantages of existing efficient algorithms such as Fast Fourier Transform (FFT) to assess dissimilarity. It is noteworthy that comparing texture windows is also used in several other applications. For example, \cite{AY+14} developed a method for a texture synthesis application where it is required to measure the dissimilarity between synthesized texture patches and the input texture to be imitated. First, features such as coherency and orientation are computed. Then, the comparison performed by~\cite{AY+14} measures the dissimilarity using various distance measures. In contrast, our method compares the texture windows directly using a very efficient dissimilarity measure without having to compute feature vectors, which can be very computationally demanding. Further, our method takes into account perceptual similarity not just objective similarity as we believe this is one of the important factors in seismic interpretation.

\subsection{Gradient of Texture}
We define the gradient of texture at a given point as the perceptual dissimilarity between two square neighborhood windows that share a side centered around the given point. The shared side is perpendicular to the direction of interest~\cite[]{GW06, PP10}.  Figure~\ref{fig:GoT} illustrates an example with two texture regions separated by a purely vertical boundary. For a given row, if we slide the center point (and hence the two neighborhood windows in the figure) along the horizontal direction, we expect the GoT profile to follow the curve shown at the bottom of the figure. For other rows, we expect similar curves because the boundary is purely vertical in this example. Theoretically, the highest dissimilarity, and hence the highest GoT value, is obtained when the center point falls exactly on the texture boundary because the left neighborhood window $W_{x-}$ will have completely different texture content from that of the right neighborhood window $W_{x+}$.  When we move the center point away from the texture boundary to the position shown in Figure~\ref{fig:GoT}, the content of $W_{x+}$ is purely from the right texture region, while $W_{x-}$ contains both textures.  The partially shared content will cause the dissimilarity value, and hence the GoT value, to drop. If we keep moving the center point along with the neighborhood windows away from the boundary, we reach a point where the contents of $W_{x-}$ and $W_{x+}$ are from the same texture region; i.e., the texture contents will be exactly similar. This exact similarity means zero dissimilarity, and hence, zero GoT.


For the texture boundary in Figure~\ref{fig:GoT}, computing the gradient of texture in the horizontal direction is sufficient because the texture boundary is purely vertical. For the general case, we compute the gradient of texture as a vector with components in both the horizontal and vertical directions. Figure~\ref{fig:GoT_xycomp} shows how the gradient of texture is computed for the general case. The solid neighborhood windows are used to compute the horizontal component, while the dashed neighborhood windows are used to compute the vertical component. Mathematically,
\begin{eqnarray}
\label{equ:got}
&&\mathbf{G_x}[i,j] = d\left(W^{i,j}_{x-}, W^{i,j}_{x+}\right),\\
&&\mathbf{G_y}[i,j] = d\left(W^{i,j}_{y-}, W^{i,j}_{y+}\right),\\
&&\mathbf{G}[i,j] = \left(\mathbf{G_x}^2[i,j]+\mathbf{G_y}^2[i,j]\right)^{\frac{1}{2}}, \label{equ:got_combined}
\end{eqnarray}
where $\mathbf{G_x}[i,j]$ and $\mathbf{G_y}[i,j]$ denote the GoT values at point $[i,j]$ along the horizontal and vertical directions respectively, $\mathbf{G}[i,j]$ represents the GoT value combining the horizontal and vertical components, function $d(\cdot)$ defines a dissimilarity measure, and the four $W^{i,j}$ expressions with four different subscripts refer to the four neighborhood windows shown in Figure~\ref{fig:GoT_xycomp}.


With the large scale variations in texture content, the neighborhood window size used to compute the GoT attribute needs to be carefully chosen. To improve robustness, we use different window sizes (5 in our case) and we choose the weighted average of the quantities as the attribute value. The neighborhood window sizes range from $3\times3$ to $11\times11$. Figure~\ref{fig:neighb_wins} shows neighborhood windows of minimum and maximum sizes around a labeled salt dome boundary. Note that different window sizes enable capturing texture dissimilarity across different scales. The multi-scale version of GoT is mathematically expressed as follows:
\begin{eqnarray}
&&\mathbf{G_x}[i,j] = \sum_{n=1}^N{w_n .d\left(W^{i,j}_{n,x-}, W^{i,j}_{n,x+}\right)},\\
&&\mathbf{G_y}[i,j] = \sum_{n=1}^N{w_n .d\left(W^{i,j}_{n,y-}, W^{i,j}_{n,y+}\right)},\\
&&\mathbf{G}[i,j] = \left(\mathbf{G_x}^2[i,j]+\mathbf{G_y}^2[i,j]\right)^{\frac{1}{2}},
\label{equ:ms-got}
\end{eqnarray}
where, in our case,  $N=5$. Also, for a given $n$, the neighborhood window size is $(2n+1)\times(2n+1)$, and $W^{i,j}_{n,x-}$, $W^{i,j}_{n,x+}$, $W^{i,j}_{n,y-}$, $W^{i,j}_{n,y+}$ respectively denote the neighborhood windows of size $(2n+1)\times(2n+1)$ that are to the left, right, bottom and top of the point $[i,j]$ where the attribute is being computed. Finally, $w_n$ denotes the weight of the GoT value computed at the window whose size is $(2n+1)\times(2n+1)$. The GoT value increases as we increase the neighborhood window size because of the dissimilarity measure we are using (as we show in the next section). Thus, with equal weights, the overall attribute would be more biased to the larger window sizes. As a result, to partially compensate for this bias, we choose $w_n$ to be inversely proportional to $n$ as $w_n=1/n$.


\subsection{Dissimilarity Measure}
\label{sec:dissim}
To measure the perceptual dissimilarity between neighboring windows, we use the following measure
\begin{equation}
d(W_{-}, W_{+}) = E\left\{\left| \mathscr{F} \left\{     \left| \mathscr{F} \left\{ \left|W_{-} - W_{+}\right| \right\} \right| \right\}  \right|\right\},
\label{eq:distanceMeasure}
\end{equation}
where $\mathscr{F}\{\cdot\}$  denotes the 2D DFT,  $E$ is the expectation operator, and $W_{-}$, $W_{+}$ denote the two neighboring windows whose dissimilarity is to be evaluated. This measure is a close variation of  the one introduced in~\cite[]{HA14globalsip}, which was shown to be consistent with human perception while being computationally efficient. The dissimilarity measure characterizes the amount of variation, or chaos, in the magnitude spectrum of the absolute difference between the two windows. The outer Fourier transform is considered the magnitude spectrum of the magnitude spectrum. Since  the proposed dissimilarity measure is highly consistent with human perception \cite[]{HA14globalsip}, we believe it partially imitates how the human interpreter delineates the salt boundaries according to the texture variations inside and outside the salt region.

\subsection{Initialization and Region Growing}

GoT detects several texture boundaries including ones outlining salt and others that are irrelevant to salt, e.g., from strong reflectors that are away from the salt dome. In order for the detection method to focus on the salt boundaries only, we identify an initialization point inside the salt dome and then use region growing until a predefined threshold for GoT magnitude is met.
The initialization point could be selected automatically or interactively with human interpreter involved.  Even with interactive selection of initialization point, the time spent by human interpreter will be considerably reduced since the interpreter would have to quickly click on \emph{any} arbitrary point within the salt region, as opposed to carefully traversing a long, tortuous  salt dome boundary. Note that the proposed method is not sensitive to the initialization point selection as long as the selected point falls anywhere within the salt region. In other words, the initialization point needs to fall within the salt region, but is otherwise arbitrary.

For automated initialization, we propose a selection method that is based on the directionality attribute~\cite[]{hegazy2014texture}. The rationale is that salt regions would have the lowest directionality in the whole seismic section under consideration. To compute the directionality map $\mathbf{D}$, we first compute the moment of inertia tensor $\mathbf{I}$ of the scattered plot of the horizontal and vertical components of the intensity gradient $\mathbf{\Delta}$ computed for a small neighborhood window $W^{i,j}$ centered around the point $[i,j]$. Note that the gradient of intensity $\mathbf{\Delta}$ used here is different from the gradient of texture (GoT) $\mathbf{G}$ introduced in the previous subsections.  Mathematically, we compute the moment of inertia tensor $\mathbf{I}$ around a given point [i,j] as follows:

\begin{equation}
\mathbf{I} = \begin{bmatrix}
I_{xx}& I_{xy}
\\
  I_{yx}& I_{yy}
\end{bmatrix},
\end{equation}

\begin{equation}
I_{xx} = \sum\limits_{[k,l]\in W^{i,j}}(\mathbf{\Delta_y}[k,l]-\bar{\Delta}_y)^2,
\end{equation}

\begin{equation}
I_{xy} = I_{yx} = \sum\limits_{[k,l]\in W^{i,j}}-(\mathbf{\Delta_x}[k,l]-\bar{\Delta}_x)(\mathbf{\Delta_y}[k,l]-\bar{\Delta}_y),
\end{equation}

\begin{equation}
I_{yy} = \sum\limits_{[k,l]\in W^{i,j}}(\mathbf{\Delta_x}[k,l]-\bar{\Delta}_x)^2,
\end{equation}

where:
\begin{itemize}
\item
$\mathbf{I}$ is the moment of inertia tensor for the scattered plot of the intensity gradient components, which is obtained for a neighborhood window $W^{i,j}$ centered around point $[i,j]$.

\item
$W^{i,j}$ is a neighborhood window centered around point $[i,j]$. The summation is performed over all the points $[k,l]$ in $W^{i,j}$.

\item
$\mathbf{\Delta_x}[k,l]$ and $\mathbf{\Delta_y}[k,l]$ respectively denote the horizontal and vertical components of the intensity gradient at a point $[k,l]$ inside $W^{i,j}$ .

\item
 $\bar{\Delta}_x$ and $\bar{\Delta}_y$ respectively denote the mean of the horizontal and vertical components of the gradient computed over the neighborhood window $W^{i,j}$.
\end{itemize}

The eigenvalues of the moment of inertia tensor correspond to the axes of minimum and maximum moments of inertia. Thus, we can use the eigenvalues of the moment of inertia tensor to capture the directionality of texture as we show below. To make the method more robust, we average the directionality maps obtained using multiple window sizes as outlined in~\cite[]{hegazy2014texture}. Let $\Lambda_{1, n}[i,j]$, $\Lambda_{2, n}[i,j]$ denote the eigenvalues of the moment of inertia tensor computed around point $[i,j]$ using a neighborhood window whose size is $(2n+1)\times(2n+1)$ and centered around [i,j]. We can mathematically define the multi-scale directionality attribute at point $[i,j]$ as:

\begin{equation}
\mathbf{D}[i,j] = \sum_{n=1}^N{1-\frac{ \min(\Lambda_{1, n}[i,j], \Lambda_{2,n}[i,j])}{  \max(\Lambda_{1, n}[i,j], \Lambda_{2,n}[i,j])}}.
\label{eq:direc}
\end{equation}

For non-salt regions, one eigenvalue will be much larger than the other eigenvalue. Thus, the directionality attribute will be close to 1. On the other hand, for salt regions, the eigenvalues will be almost equal, leading to directionality attributes close to zero.

For added reliability and to protect against selecting a noisy point that happened to have very low directionality, we smooth out the directionality map with a Gaussian filter. We obtain the initialization point by minimizing the smoothed multi-scale directionality map. Therefore, the initialization point is selected according to:

\begin{equation}
\left\{
\begin{aligned}
&\arg \min_{[i,j]} {\mathbf{D} \ast \mathbf{\Omega}}\\
&\mathbf{\Omega}[u,v]=e^{-\frac{d_{\mathbf{\Omega}}^2(u,v)}{2\sigma_{\mathbf{\Omega}}^2}},\quad u, v\in\left\{1,2,\cdots, 2N+1\right\}
\end{aligned}
\right.
\end{equation}%
where $\mathbf{\Omega}$ is a Gaussian kernel with a size of $(2N+1)\times (2N+1)$ used to smooth out the multi-scale directionality attribute map and $\ast$ is the 2D convolution operator. The elements in $\mathbf{\Omega}$ indexed by $u$ and $v$ are determined by function $d_{\mathbf{\Omega}}$, which calculates the distance from $[u,v]$ to the center of the Gaussian kernel. In addition, $\sigma_{\mathbf{\Omega}}$ denotes the standard deviation of the Gaussian kernel.

We grow the initialization point into a region. Since the initialization point is inside the salt dome, i.e., inside a texture region, we expect the texture gradient around the initialization point to be very low (theoretically zero). Thus, the initialization point can be grown since GoT in the neighboring region is less than a certain threshold value (determined according to the next subsection). As we grow the region, we start meeting points that are close to the boundary of the dome where the GoT value is high. We stop growing along those points that reached the threshold of the GoT map. In the following subsection, we discuss how the threshold value is determined.

\subsection{Thresholding}

The threshold value can be determined interactively with a human interpreter or automatically. For automated threshold value selection, we propose using Otsu's method \cite[]{O79}. The algorithm assumes that there are two classes of pixels following bimodal histogram. The optimal threshold is therefore calculated as the one that minimizes the intra-class variance, which is the weighted sum of the variances of the two classes. Thus, the optimal threshold is given by:

\begin{equation}
\arg \min_T \left\{ \sigma_1^2(T) \sum_{i=0}^{T-1}{p(i)} + \sigma_2^2(T) \sum_{i=T}^{N-1}{p(i)} \right\},
\end{equation}
where $T$ is the threshold value, $p(i)$ is the probability of the intensity value $i$, and $\sigma_1^2(T) $, $\sigma_2^2(T) $ are the variances of the first and second classes, respectively. The first and second classes are determined given the threshold $T$. The optimal threshold value is found by exhaustively searching for $T$ between $0$ and $N-1$.

For more efficient implementation, Otsu shows that minimizing the intra-class variance is equivalent to maximizing the inter-class variance, which can be simplified as follows:
\begin{equation}
\arg \max_T \left\{ \left(\sum_{i=0}^{T-1}{p(i)}\right)     \left(\sum_{i=0}^{T-1}{p(i)} \right)       \left(\mu_1(T) - \mu_2(T)\right) \right\} ,
\end{equation}
where $\mu_1(T)$ and $\mu_2(T)$ are the mean values of the first and second classes, respectively given a threshold value $T$ \cite[]{O79}.

\subsubsection{Noise Impact}
In this subsection, we discuss the impact of acquisition noise on the salt boundaries detected on the migrated seismic data. \cite{BT08} study the impact of shot gathers contaminated with Gaussian noise on the migrated seismic data. Under Gaussian noise, the reflectors in the migrated image are correctly positioned but are affected by noise \cite[]{BT08}. In the following analysis, we assume that the random acquisition noise will result in a migrated image that is contaminated with additive random noise.

Let $\hat{W}_1$, $\hat{W}_2$ be two noise-contaminated adjacent neighborhood windows whose dissimilarity is to be evaluated to compute the GoT attribute at their center point, such that
\begin{eqnarray}
&&\hat{W}_1= W_1+n_1,\\
&&\hat{W}_2 = W_2+n_2,
\end{eqnarray}
where $W_1$, $W_2$ denote the noise-free windows, and  $n_1$, $n_2$ denote random noise added to  $W_1$ and $W_2$, respectively. We re-write the dissimilarity measure in  Eq.(\ref{eq:distanceMeasure})  between the noise contaminated windows as follows:
\begin{equation}
d(\hat{W}_1, \hat{W}_2) = E\left\{\left| \mathscr{F} \left\{     \left| \mathscr{F} \left\{ \Delta_W + \Delta_n\right\} \right| \right\}  \right|\right\}.
\end{equation}

For simplicity, we remove the absolute value of the input to the inner 2D DFT as we noticed through experimentation that it does not make any significant difference in the performance of the proposed method. 

Using the linearity property, we can express the inner 2D DFT as follows
\begin{equation}
\mathscr{F} \left\{ \Delta_W + \Delta_n\right\}  = \mathscr{F} \left\{ \Delta_W \right\}   + \mathscr{F} \left\{ \Delta_n\right\}
\end{equation}

Thus, the 2D DFT coefficients are the same as those of the noise-free case $\mathscr{F} \left\{ \Delta_W \right\}$, except they are contaminated with the noise signal $\mathscr{F} \left\{ \Delta_n\right\}$. According to~\cite{SR86}, we can express this noise signal as

\begin{equation}
\mathscr{F} \left\{ \Delta_n\right\}  = \sum_{k=0}^{2n} \sum_{l=0}^{2n}  a[k,l]+ j\sum_{k=0}^{2n} \sum_{l=0}^{2n}  b[k,l],
\end{equation}
where each summation is performed over the square neighborhood window size of $(2n+1)\times(2n+1)$, and $a$ and $b$ are random variables. The noise signals affecting the real and imaginary parts of Fourier coefficients are derived from the sum of a number of random variables. Therefore, the original Fourier coefficients are subjected to additive random noise. According to~\cite{SR86}, the covariance matrix of the real and imaginary parts is nearly diagonal, which means that the real and imaginary coefficient noise components are independent.

Both $\mathscr{F} \left\{ \Delta_W \right\}$ and $\mathscr{F} \left\{ \Delta_n\right\}$ are complex. As a result, the absolute value of $\mathscr{F} \left\{ \Delta_W \right\}   + \mathscr{F} \left\{ \Delta_n\right\}$ can be determined by vector algebra as shown in Figure~\ref{fig:sn_vectors}, where $\vec{s}$ refers to $\mathscr{F} \left\{ \Delta_W \right\}$ and $\vec{n}$ refers to $\mathscr{F} \left\{ \Delta_n \right\}$. If the signal to noise ratio is much larger than 1, then the angle $\theta$ in Figure~\ref{fig:sn_vectors} is very small. As a result, noise impact on the absolute value of Fourier coefficients can be approximated by simply projecting the noise component $\vec{n}$ on the original signal $\vec{s}$ vector direction. Consequently, such noise can be approximated as additive random noise $n_m$ as follows.

\begin{equation}
n_m = |\mathscr{F} \left\{ \Delta_n\right\}| \cos \phi,
\end{equation}
where $\phi$, which is a random variable, denotes the angle between the signal and noise vectors in Figure~\ref{fig:sn_vectors}.




Consequently, $|\mathscr{F} \left\{ \Delta_W \right\}   + \mathscr{F} \left\{ \Delta_n\right\}|$, which is the input to the outer 2D DFT, suffers additive random noise $n_m$.  Similarly, applying the same derivation for the magnitude of the outer 2D DFT, we conclude that the effect of noise on the magnitude of the outer 2D DFT coefficients suffer from additive random noise as well, say $n_d$. Since the dissimilarity measure $d()$ is the mean value of the the outer 2D DFT magnitude, then $d()$ is shifted by the mean of $n_d$, that is $\mu_d$.

Theoretically, the additive random noise affecting the migrated images shifts the dissimilarity measure (and hence, GoT) by a constant value $\mu_d$, assuming the noise is stationary. Because we apply an adaptive threshold to GoT, we expect almost no effect on the detection results. Practically, however, the window sizes are finite, and therefore, the proposed method will still suffer under noise because the averaging would be performed for a few samples. We conclude that to make the method more robust under noise, it is important to drop the smaller window sizes and use larger window sizes.

\subsection{Salt Dome Tracking}
\label{sec:tracking}
In the detection of salt dome boundaries, the accuracy of labeled boundaries depends greatly on the selection of certain parameters. Since salt dome structures commonly vary across seismic sections, interpreters have to adjust parameters so that detected boundaries in each section can accurately capture structural changes. However, extra labor and time involved in the tuning of parameters may lower interpretation efficiency. Therefore, to avoid the frequent adjustment of parameters, we group seismic sections into reference and predicted sections by borrowing the ideas of I- and P-frames from video-coding techniques and apply the proposed detection method only to reference sections that constitute the minority of seismic sections. By tracking detected boundaries in reference sections through a seismic volume, we can synthesize salt dome boundaries in predicted sections with limited human intervention. The lower part of Figure~\ref{fig:diagram} illustrates the pipeline of the proposed tracking method, each step of which is going to be introduced in detail in the following subsections.

\subsubsection{Tensors and Multi-linear Analysis}
\label{sssec:tensor}
By observing salt domes in seismic sections as Figure~\ref{fig:SaltDome} shows, we notice that local areas along salt dome boundaries commonly have similar textures. Therefore, by classifying boundary textures in the reference section, we can build texture tensors and employ the tensor-based subspace learning algorithm to acquire their texture features, which act as important constraints to the synthesis of salt dome boundaries in predicted sections. Before we introduce the proposed classification method, we briefly review the basic concepts of tensors and multi-linear analysis.

In multi-linear algebra, a tensor represents a multidimensional array. Conventionally, we denote tensors using calligraphic letters such as $\mathcal{A}$, and matrices using bold uppercase letters such as $\mathbf{U}$. An $N$th-order tensor $\mathcal{A}\in\mathbb{R}^{I_1\times I_2\times\cdots\times I_N}$ with entries denoted $\mathcal{A}(i_1i_2\cdots i_N)$ has $N$ indices or modes. By unfolding tensor $\mathcal{A}$ along the $n$th mode, we obtain matrix $\mathbf{A}^{(n)}\in\mathbb{R}^{I_n\times(I_1\times\cdots\times I_{n-1}\times I_{n+1}\cdots\times I_N)}$. Figure~\ref{fig:unfolding} illustrates the unfolding of a third-order tensor along three modes, distinguished by specific colors.
%
%
The inverse operation of n-mode unfolding is n-mode folding, which restores $\mathcal{A}$ from matrix $\mathbf{A}^{(n)}$. The n-mode product of tensor $\mathcal{A}$ by matrix $\mathbf{U}\in\mathbb{R}^{J_n\times I_n}$, denoted $\mathcal{A}\times_n\mathbf{U}$, defines new tensor $\mathcal{B}$ with entries calculated as
\begin{equation}
\label{equ:tensorMulti}
\Scale[0.96]{
\mathcal{B}(i_1\cdots i_{n-1}j_ni_{n+1}\cdots i_N)=\sum\limits_{i_n}\mathcal{A}(i_1\cdots i_N)\cdot \mathbf{U}(j_n,i_n)}.
\end{equation}
In addition, the product above can also be implemented by folding $\mathbf{U}\cdot\mathbf{A}^{(n)}$ along the $n$th mode. Similar to the singular value decomposition (SVD) of matrices, the decomposition of tensors can be expressed as $\mathcal{A}=\mathcal{S}\times_1\mathbf{U}^{(1)}\times_2\mathbf{U}^{(2)}\cdots\times_N\mathbf{U}^{(N)}$, in which $\mathbf{U}^{(n)}$, $n=1,2,\cdots,N$, represents an $I_n\times I_n$ orthogonal matrix with column vectors spanning the column space of $n$-mode unfolding matrix $\mathbf{A}^{(n)}$~\cite[]{de2000best}. Since a tensor with high dimensions requires compact representation in real applications, we can apply multi-linear principal component analysis (MPCA)~\cite[]{lu2008mpca} to extract a low-dimensional tensor subspace that captures the most variance of the original tensor. Such a tensor subspace, denoted $\mathcal{C}\in\mathbb{R}^{P_1\times P_2\times\cdots\times P_N}$, $P_n<I_n$, $n=1,2,\cdots,N$, can be obtained by projecting the original tensor $\mathcal{A}$ onto matrices $\widetilde{\mathbf{U}}^{(n)}$ as
\begin{equation}
\label{equ:mpca}
\mathcal{C}=\mathcal{A}\times_1\widetilde{\mathbf{U}}^{(1)^T}\times_2\widetilde{\mathbf{U}}^{(2)^T}\cdots\times_N\widetilde{\mathbf{U}}^{(N)^T},
\end{equation}
where $\widetilde{\mathbf{U}}^{(n)}\in\mathbb{R}^{I_n\times P_n}$ is composed of eigenvectors corresponding to the largest $P_n$ eigenvalues of $\mathbf{A}^{(n)}\cdot\mathbf{A}^{(n)^T}$.

\subsubsection{Texture Tensor Classification}
\label{ssec:textureTensor}

To extract texture features from salt dome boundaries detected in a reference section, we define pairs of patches centered at boundary points that are subimages containing the boundary textures of the reference section and its corresponding GoT map. Since matrices are particular third-order tensors with a third-mode dimension equal to one, we denote patch pairs in a form of tensor pairs as $\{\mathcal{S}_{p_i}, \mathcal{G}_{p_i}\}\subset\mathbb{R}^{I_1\times I_2\times1}$, $i=1,2,\cdots,N_b$, where $N_b$ represents the number of all boundary points and $I_1$ and $I_2$ define the dimensions of patches along the depth and crossline directions, respectively. Since textures have no bias on either the depth or crossline direction, we commonly assume that $I_1$ is equal to $I_2$. In addition, to ensure that boundary textures can be captured by square-shaped patches, we determine the edge length of patches as around one-tenth of the larger of the vertical and horizontal resolutions. Furthermore, by evaluating the similarity between patches, we can classify boundary textures and build a set of texture tensors, denoted $\{\mathcal{S}_k, \mathcal{G}_k\}$, $k=1,2,\cdots,N_t$.

\begin{figure*}[b]
\centering
\begin{minipage}[b]{0.2\linewidth}
  \centering
  \centerline{\includegraphics[height=3cm]{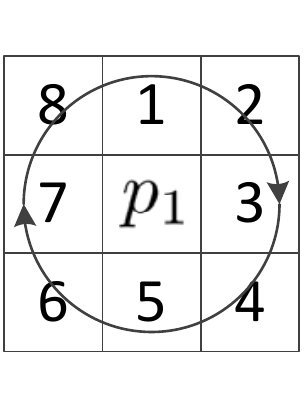}}
  \centerline{(a)}\medskip
\end{minipage}
\begin{minipage}[b]{0.7\linewidth}
  \centering
  \centerline{\includegraphics[width=7cm]{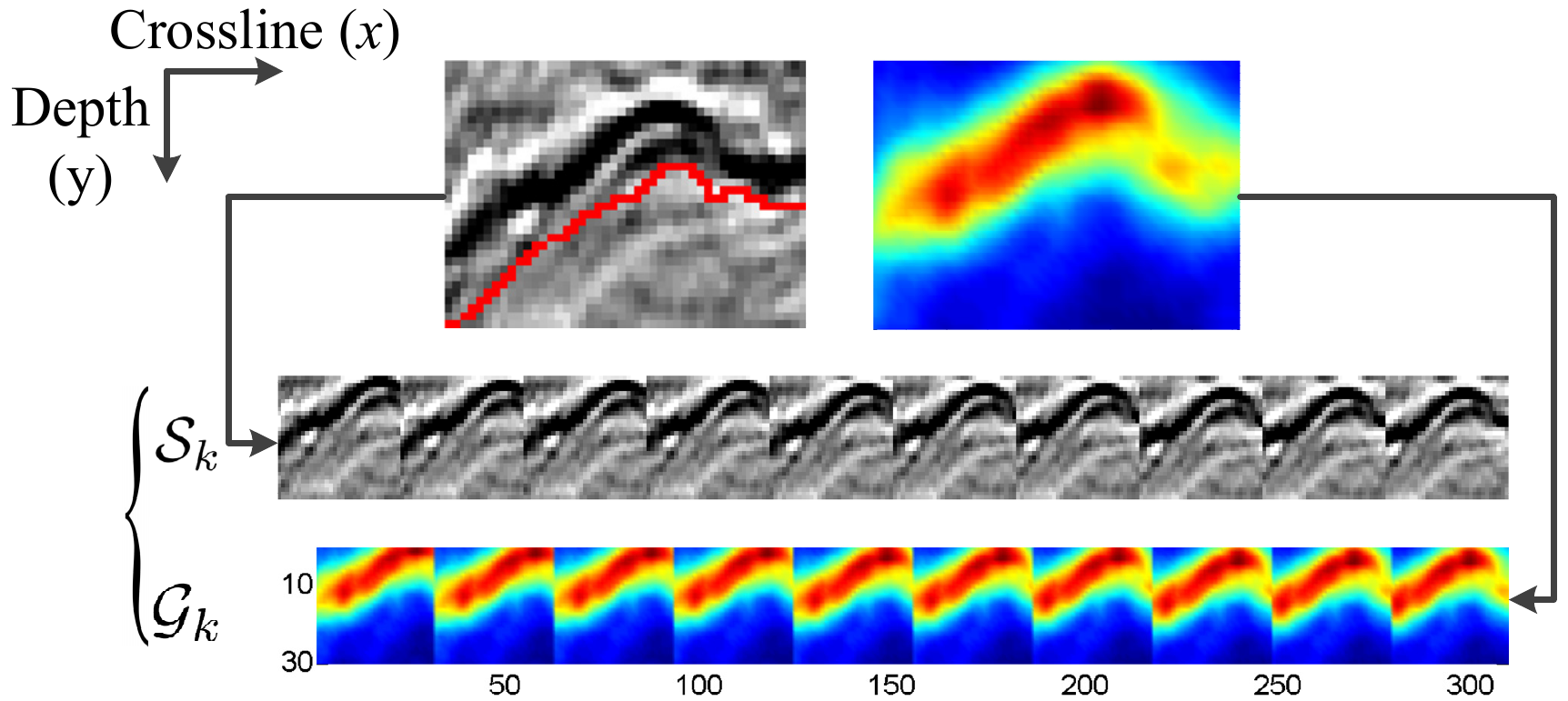}}
  \centerline{(b)}\medskip
\end{minipage}
\caption{(a) The priority of points in the $3\times 3$ neighborhood of $p_1$ and (b) The 3-mode unfolding matrices of a tensor pair extracted from the local area of a seismic section and its corresponding GoT map.}
\label{fig:classified}
\end{figure*}

To clarify the iterative classification strategy, we depict its block diagram in Figure~\ref{fig:classification}.
Because of the roughly semi-circular shape of detected boundaries in the reference section, we traverse clockwise along the boundary points and define $p_1$ as the first point in the bottom-left corner satisfying the dimension constraint of patches. After initializing $\{\mathcal{S}_1,\mathcal{G}_1\}$ with patch pair $\{\mathcal{S}_{p_1},\mathcal{G}_{p_1}\}$, we search for $p_2$ in the $3\times 3$ neighborhood of $p_1$ on the basis of the point priority shown in Figure~\ref{fig:classified}(a), where a smaller number represents a higher priority. To facilitate the evaluation of the texture similarity between the patch pair of $p_2$, $\{\mathcal{S}_{p_2},\mathcal{G}_{p_2}\}$, and the initial tensor pair, $\{\mathcal{S}_1,\mathcal{G}_1\}$, we append $\mathcal{S}_{p_2}$ and $\mathcal{G}_{p_2}$ to $\mathcal{S}_{1}$ and $\mathcal{G}_{1}$, respectively, along the third mode and obtain the updated tensor pair $\{\widetilde{\mathcal{S}}_{1},\widetilde{\mathcal{G}}_{1}\}$. In Figure~\ref{fig:classification}, function $(\cdot|\cdot)$ describes an operation that extends tensors along the third direction. Furthermore, by applying the MPCA to $\{\widetilde{\mathcal{S}}_{1},\widetilde{\mathcal{G}}_1\}$, we obtain the projection matrix of each mode, denoted $\widetilde{\mathbf{U}}_{\widetilde{\mathcal{M}}_1}^{(n)}\in\mathbb{R}^{I_n\times P_n}$, $(n=1,2)$, $\widetilde{\mathbf{U}}_{\widetilde{\mathcal{M}}_1}^{(3)}\in\mathbb{R}^{(I_1\times I_2)\times P_n}$,  $\mathcal{M}=\{\mathcal{S},\mathcal{G}\}$, in which $[P_1,P_2,P_3]$ are determined empirically. Because of the high computational efficiency of the sequential Karhunen-Loeve (SKL) algorithm~\cite[]{levey2000sequential,hu2011incremental} on updating projection matrices, we extract $\widetilde{\mathbf{U}}_{\widetilde{\mathcal{M}}_1}^{(3)}$, $\mathcal{M}=\{\mathcal{S},\mathcal{G}\}$ from the row spaces of 3-mode unfolding matrices instead of their column spaces. Since eigenvectors in the projection matrices reflect the texture features of tensors, to evaluate the similarity between patches and tensors, we define reconstruction error $e_c$ as follows:
\begin{equation}
\label{equ:reconError1}
\Scale[0.85]{
\begin{aligned}
e_c=&\sum\limits_{n=\{1,2\},\mathcal{M}=\{\mathcal{S},\mathcal{G}\}}\left\|\mathcal{M}_{p_i}-\mathcal{M}_{p_i}\times_n\left(\widetilde{\mathbf{U}}^{(n)}_{\widetilde{\mathcal{M}}_k}\cdot\widetilde{\mathbf{U}}^{(n)^{T}}_{\widetilde{\mathcal{M}}_k}\right)\right\|_{F}^2+\\
&\sum\limits_{\mathcal{M}=\{\mathcal{S},\mathcal{G}\}}\left\|\mathbf{M}_{p_i}^{(3)}-\mathbf{M}_{p_i}^{(3)}\cdot\widetilde{\mathbf{U}}_{\widetilde{\mathcal{M}}_k}^{(3)}\cdot\widetilde{\mathbf{U}}_{\widetilde{\mathcal{M}}_k}^{(3)^T}\right\|_F^2   \end{aligned}},
\end{equation}
where $p_i$ and $k$ represent the indices of the current boundary point and tensor pair, respectively. If $e_c$ is less than threshold $T_e$, the strong similarity allows us to extend $\{\mathcal{S}_1,\mathcal{G}_1\}$ by appending $\{\mathcal{S}_{p_2},\mathcal{G}_{p_2}\}$. Otherwise, the great deviation requires us to keep $\{\mathcal{S}_1,\mathcal{G}_1\}$ unchanged and initialize another tensor pair $\{\mathcal{S}_2,\mathcal{G}_2\}$ with $\{\mathcal{S}_{p_2},\mathcal{G}_{p_2}\}$. By repeating the step above, we obtain classified tensor pairs that contain the local texture information of the detected boundary. Figure~\ref{fig:classified}(b) illustrates one local area of a seismic section and its corresponding GoT map, in which the red curve represents the labeled salt dome boundary. In addition, Figure~\ref{fig:classified}(b) also shows the 3-mode unfolding matrices of a tensor pair extracted from this local area, in which texture patches have a strong correlation.
%

\subsubsection{Tracked Boundary Synthesis}
\label{ssec:boundarySynthesis}

On the basis of texture features extracted from classified texture tensors, we can localize boundary points in predicted sections. We first project the detected boundary in the reference section onto the target section and keep the coordinates of all points unchanged. Then, to identify the tracked point of every projected point $p$, we search among candidate points located along the normal direction of the projected boundary within radius $R_s$, which is adaptively determined by the inline number difference between the reference and the current predicted sections. To define the textures of candidate points, we extract pairs of patches centered at these points from the predicted section and its corresponding GoT map, denoted $\{\mathcal{S}_{p,j}, \mathcal{G}_{p,j}\}\subset\mathbb{R}^{I_1\times I_2\times 1}$, $j=1,2,\cdots,(2R_s+1)$, where $j$ represents the index of the candidate point. Since in the reference section the patch pair of projected point $p$ belongs to tensor pair $\{\mathcal{S},\mathcal{G}\}$,
by comparing the similarity between $\{\mathcal{S}_{p,j}, \mathcal{G}_{p,j}\}$ and $\{\mathcal{S},\mathcal{G}\}$, we can determine tracked point $p^{\ast}$. The block diagram of the proposed tracking process is shown in Figure~\ref{fig:localization}.
The reconstruction error $e_t$ is calculated as follows:
\begin{equation}
\label{equ:reconError2}
\Scale[0.77]{
        \begin{aligned}
        e_t=&\sum\limits_{n=\{1,2\},\mathcal{M}=\{\mathcal{S},\mathcal{G}\}}\lambda_{\mathcal{M}}\cdot\left\|\mathcal{M}_{p,j}-\mathcal{M}_{p,j}\times_n\left(\widetilde{\mathbf{U}}^{(n)}_{\widetilde{\mathcal{M}}}\cdot\widetilde{\mathbf{U}}^{(n)^{T}}_{\widetilde{\mathcal{M}}}\right)\right\|_{F}^2+\\
        &\sum\limits_{\mathcal{M}=\{\mathcal{S},\mathcal{G}\}}\lambda_{\mathcal{M}}\cdot\left\|\mathbf{M}_{p,j}^{(3)}-\mathbf{M}_{p,j}^{(3)}\cdot\widetilde{\mathbf{U}}_{\widetilde{\mathcal{M}}}^{(3)}\cdot\widetilde{\mathbf{U}}_{\widetilde{\mathcal{M}}}^{(3)^T}\right\|_F^2 \end{aligned},}
\end{equation}
where $\lambda_{\mathcal{S}}=1$ and $\lambda_{\mathcal{G}}=|\log\left(\mathbf{G}[i,j]\right)|$ represent the weights of texture features extracted from seismic sections and corresponding GoT maps, respectively. $[i,j]$ denote the coordinates of the candidate point. Since GoT maps have been normalized to a range of zero to one, a larger $\mathbf{G}[i,j]$ indicates a smaller weight, which can move tracked points towards the real salt-dome boundaries of predicted sections by lowering the reconstruction error. Finally, the candidate point with the smallest reconstruction error is selected as the tracked point. We apply the same process to all boundary points and obtain binary image $\mathbf{B}$, which contains initial tracked points.

Because of the complicated structures of salt domes, it is inevitable that tracked points involve some outliers that may harm the smoothness and continuity of tracked boundaries. To remove these outliers, we apply the median filter to $\mathbf{B}$ and obtain filtered image $\mathbf{\widehat{B}}$ as follows:
\begin{equation}
\label{equ:median}
\Scale[0.79]{
\mathbf{\widehat{B}}[i,j]=\left\{
\begin{aligned}
&0\qquad\qquad\qquad\quad\  \mathbf{B}[i,j]-\underset{R_m\times R_m}{\mbox{median}}(\mathbf{B}[i,j])=-1\\
&\underset{R_m\times R_m}{\mbox{median}}(\mathbf{B}[i,j])\quad \mbox{Otherwise}
\end{aligned}
\right.},
\end{equation}
where function $\underset{R_m\times R_m}{\mbox{median}(\cdot)}$ applies the median filter to the $R_m\times R_m$ analysis window of point $[i,j]$. Without involving new points generated by median filtering,  Equation~(\ref{equ:median}) removes outliers and ensures the accuracy of the remaining tracked points. Furthermore, we connect the remaining tracked points clockwise with straight lines and obtain the initial labeling of the salt dome boundary. Since the connection based on lines may result in a jagged shape of the labeled boundary, which is geologically unreasonable, to delineate the salt-dome boundary more accurately, we need to employ two post-processing steps, dilation and skeletonization. We dilate initially labeled result $\mathbf{L}$ using disk-shaped structural element $\mathbf{H}$ as follows:
\begin{equation}
\label{equ:dilate}
\quad\mathbf{L}\oplus\mathbf{H}=\underset{z\in\mathbf{L}}{\bigcup}\mathbf{H}_z,
\end{equation}
where $\mathbf{H}_z$ represents the structural element centered at 2D point $z$.
The dilated region can be understood as the locus of the points covered by $\mathbf{H}$ when the center of $\mathbf{H}$ moves inside $\mathbf{L}$. Furthermore, by applying skeletonization~\cite[]{kong1996topological} to the dilated region, we can extract the tracked boundary.

\subsubsection{Salt Dome Tracking based on Noise-adjusted MPCA}
In our previous discussion, we commonly assume that noise does not exist in seismic datasets. However, in real cases, it is possible that migrated seismic sections contain various levels of noise. If we still depend on the MPCA-based tracking method, extracted features corresponding to the most variations of texture tensors may have low signal-to-noise ratios (SNRs). Therefore, the appearance of noise may reduce the accuracy of tracked boundaries. To deal with this problem, on the basis of Roger's work~\cite[]{roger1994faster}, we modify the proposed tracking strategy and introduce noise-adjusted MPCA (NA-MPCA) that ranks principal components (PCs) along different modes by the SNR rather than the variation. Since noise is commonly assumed to be additive, we define image patches extracted from seismic sections as $\mathbf{X}=\mathbf{S}+\mathbf{N}$, where $\mathbf{S}$ is the image patch without noise and $\mathbf{N}$ represents noise that are independent to $\mathbf{S}$. In this paper, we estimate the amplitude of noise using wavelet-based denoising method~\cite[]{chang2000adaptive}. In addition, we denote the covariance matrices of $\mathbf{X}$, $\mathbf{S}$, and $\mathbf{N}$ as $\mathbf{\Sigma}_{\mathbf{X}}$, $\mathbf{\Sigma}_{\mathbf{S}}$, and $\mathbf{\Sigma}_{\mathbf{N}}$, respectively. To arrange PCs in the order of SNRs, we need to apply a noise-whitening process that transforms $\mathbf{N}$ to identity covariance matrix $\mathbf{\widehat{N}}$. Therefore, we define transform matrix $\mathbf{P}$ and analyze the noise-whitening process as follows:
\begin{equation}
\label{equ:whitening}
\mathbf{\Sigma_{\widehat{\mathbf{N}}}}=\mathbf{P}^{T}\mathbf{\Sigma}_{\mathbf{N}}\mathbf{P}=\mathbf{P}^{T}\mathbf{V}\mathbf{\Delta}_{\mathbf{N}}\mathbf{V}^{T}\mathbf{P}=\mathbf{I},
\end{equation}
where orthonormal matrix $\mathbf{V}$ contains the eigenvectors of $\mathbf{\Sigma}_{\mathbf{N}}$. To satisfy the constraint in Equation~(\ref{equ:whitening}), we yield transform matrix $\mathbf{P}=\mathbf{V}\mathbf{\Delta}_{\mathbf{N}}^{-1/2}$. Then, we implement the noise-whitening process in the proposed salt dome tracking method by replacing the image patch $\mathcal{S}_{p_i}$, $i=1,2,\cdots,N_b$, with $\mathcal{S}_{p_i}\times_1 \mathbf{P}$ and keeping $\mathcal{G}_{p_i}$ unchanged because of the blurring effect of noise to GoT maps. The NA-MPCA removes the influence of different noise levels in the reference and predicted sections and ensures the robustness of the proposed tracking algorithm.   

\subsection{Similarity Index Measurement of Salt Dome Boundaries}
\label{ssec:index}
By employing the proposed salt dome detection and tracking method, we can semi-automatically delineate salt-dome boundaries in seismic sections. To evaluate the performance of the proposed method, we introduce a salt dome similarity (SalSIM) index that measures the distances between semi-automatically labeled results and the ground truth provided by experienced interpreters. To explain the definition of the SalSIM index, in Figure~\ref{fig:localDist} we draw  two types of curves for ilustration. The proposed SalSIM index is developed based on the Fr\'echet distance~\cite[]{alt1995computing}, which can more accurately measure the deviation between two curves than the Haudorff distance~\cite[]{huttenlocher1993comparing} by taking the continuity of curves into account.

On the basis of the Fr\'echet distance, we attempt to numerically describe local and global deviations between labeled results and the ground truth using local and global items in the proposed SalSIM index, respectively. To compare the local details of labeled boundaries and the ground truth, we define an analysis window that identifies a pair of curve segments, as the red rectangular shown in Figure~\ref{fig:localDist}. Therefore, by moving the analysis window along the ground truth and calculating the Fr\'echet distance of every pair of local segments, we can obtain a sequence of distances, $\mathbf{d}=\left[d_{i}\right]$, $i=1,2,\cdots,N_d$, where $N_d$ is the total number of pairs. The mean and standard deviation of $\mathbf{d}$, denoted $\mu_d$ and $\sigma_d$, respectively, constitute the local item of the proposed SalSIM index. In addition, the global item involves the Fr\'echet distance of two entire curves, denoted $d_{max}$, which equals the largest distance in $\mathbf{d}$. Therefore, on the basis of obtained statistics, we introduce the SalSIM index as
\begin{equation}
\label{equ:SalSIM}
\mbox{SalSIM}=\underbrace{e^{-\alpha\cdot(\mu_d+\sigma_d)}}_\text{\footnotesize{Local Item}}\cdot\underbrace{e^{-\beta\cdot d_{max}}}_\text{\footnotesize{Global Item}},
\end{equation}
where $\alpha$ and $\beta$ are normalization factors determined empirically. Since the exponential function defined on negative real numbers has a range of zero to one, we apply it in the SalSIM index for normalization. Therefore, according to the expression of SalSIM in Equation~(\ref{equ:SalSIM}), a greater SalSIM value represents smaller deviation between the semi-automatically labeled salt dome boundary and the ground truth and vice versa.

\section{Examples}
In this paper, we evaluate the performance of our proposed detection and tracking method on a real seismic volume acquired from the Netherlands offshore F3 block with the size of $24\times16\mbox{ km}^2$ in the North Sea~\cite[]{f3opendtect}. We focus on a local volume containing a distinguishable salt dome that has the inline number ranging from \#389 to \#409, the crossline number ranging from \#401 to \#701, and the time interval between 1,300ms and 1,848ms sampled every 4ms. Figure~\ref{fig:Inline400} illustrates a seismic section, Inline \#400, extracted from the local volume.

\subsection{Salt Dome Detection}
\label{ssec:expriDetect}

The proposed salt dome detection method delineates the boundaries of salt domes based on the GoT maps of seismic sections, in which the point value represents the dissimilarity of neighboring square windows. To capture the multi-scale texture contrast of neighborhoods, we calculate GoT values by varying the window size from $3\times 3$ to $11\times 11$ and averaging the corresponding dissimilarity measures, denoted $\mathbf{G}$ in Equation~(\ref{equ:got_combined}). Figure~\ref{fig:detection}(a) shows the GoT map of Inline \#400, in which likely boundary regions have higher GoT values. On the basis of GoT maps, we utilize the region-growing method to identify salt dome areas and their corresponding boundaries. In the growing process, we tweak threshold $T_g$ for each seismic section to yield the best growing result. In Figure~\ref{fig:detection}(b), we compare the green detected boundary with the manually labeled red ground truth and notice that the former overlaps the most part of the latter except the bottom-left and -right corners.
%
%

\subsubsection{Subjective Comparison of Detected Boundaries}
\label{ssec:detection_subjective}

\begin{figure*}[b]
\centering
\begin{minipage}[b]{0.5\linewidth}
  \centering
  \centerline{\includegraphics[height=3.3cm]{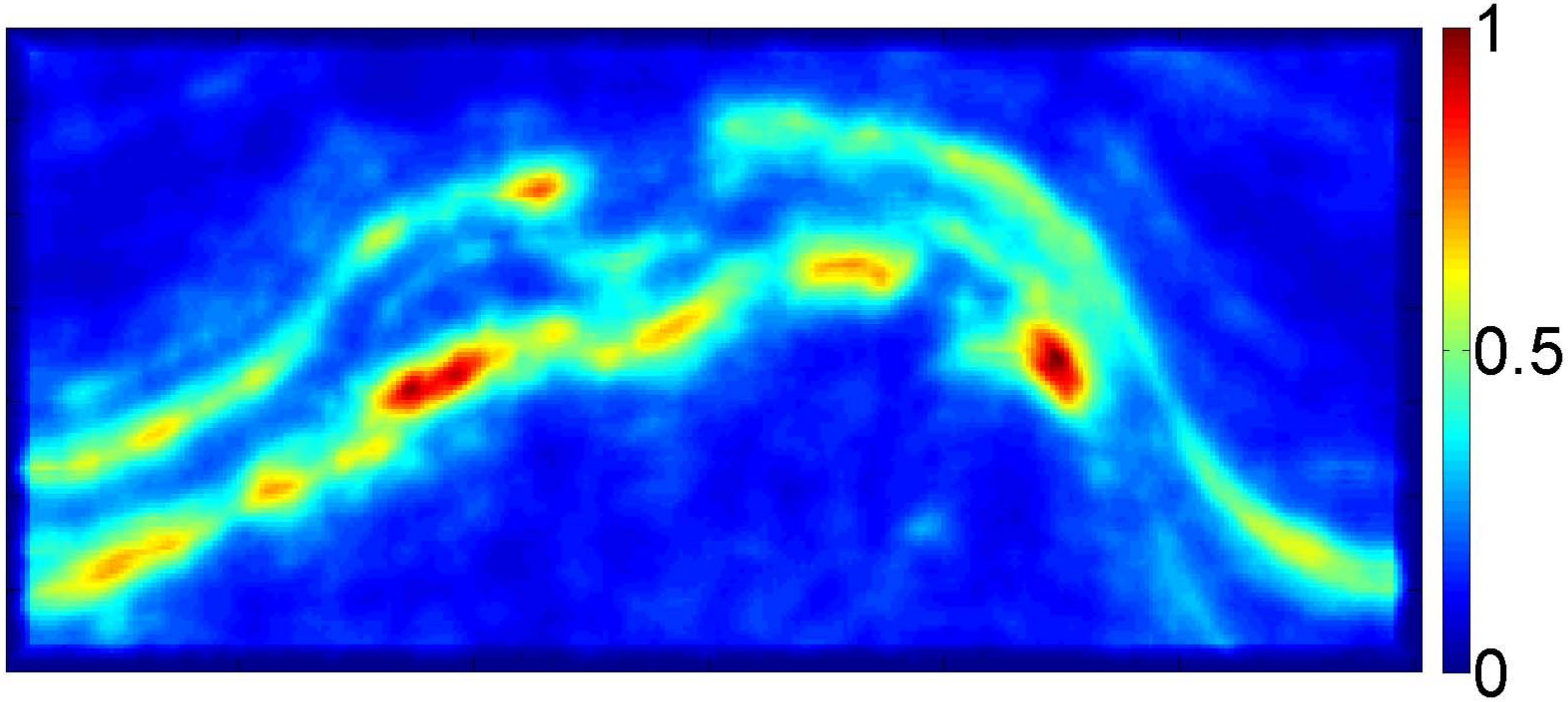}}
  \centerline{(a)}\medskip
\end{minipage}
\begin{minipage}[b]{0.48\linewidth}
  \centering
  \centerline{\includegraphics[height=3.3cm]{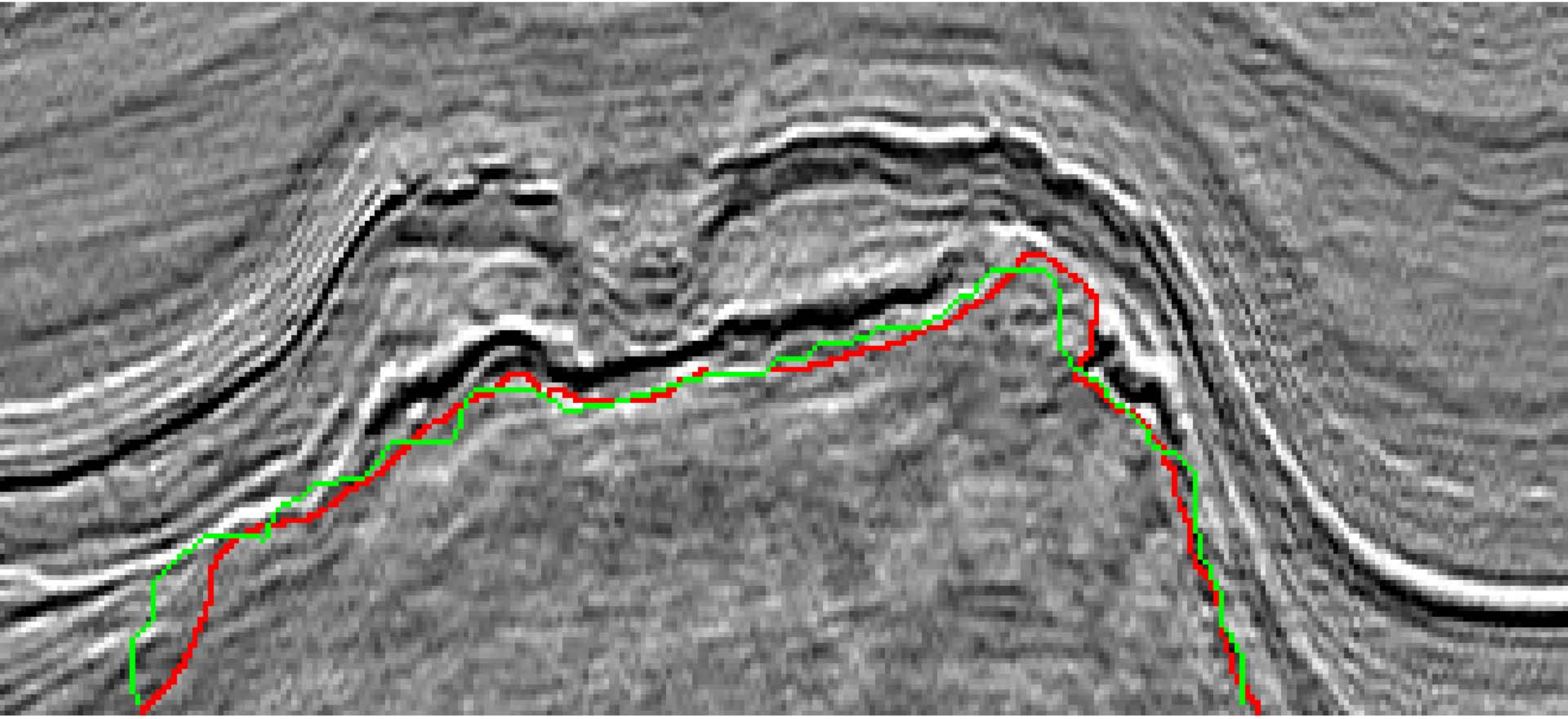}}
  \centerline{(b)}\medskip
\end{minipage}
\begin{minipage}[b]{0.5\linewidth}
  \centering
  \centerline{\includegraphics[height=3.3cm]{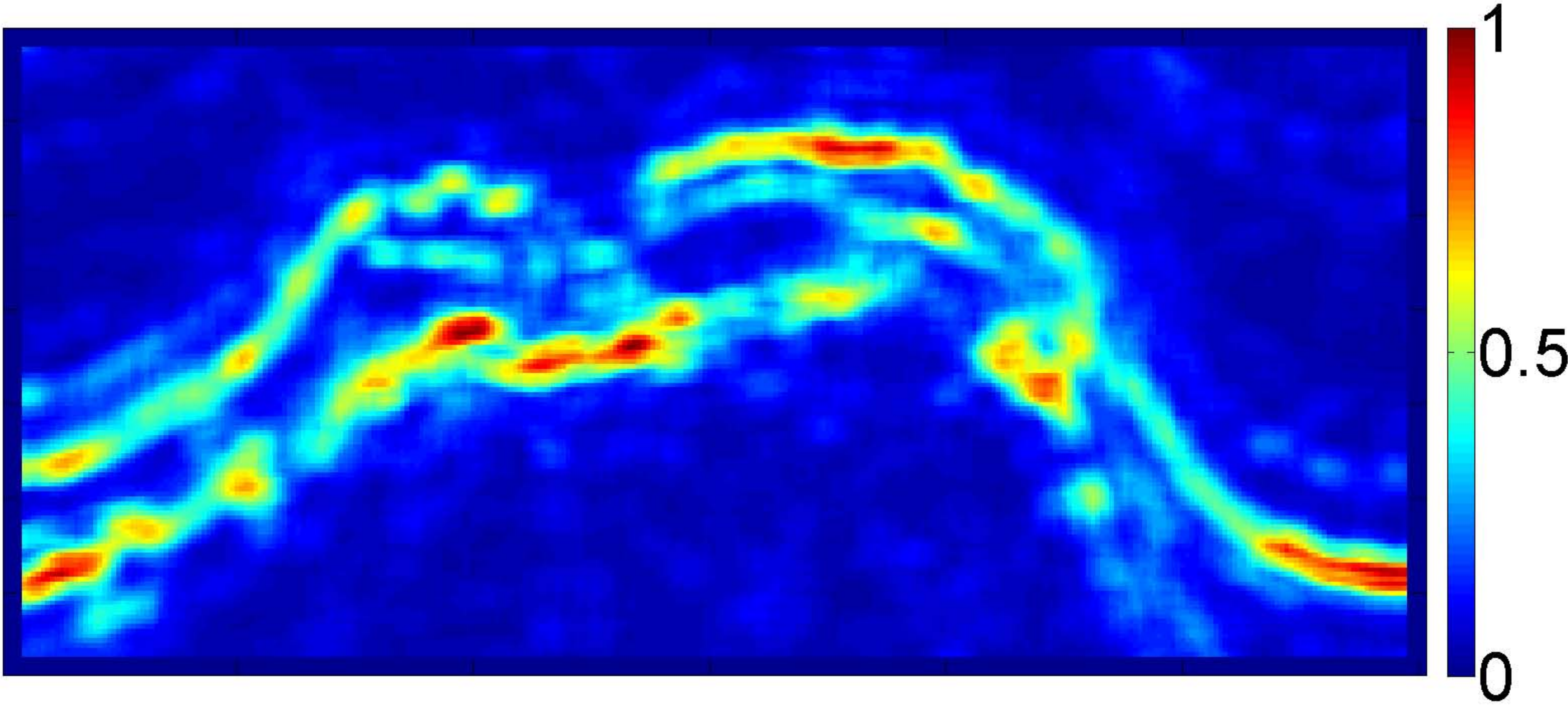}}
  \centerline{(c)}\medskip
\end{minipage}
\begin{minipage}[b]{0.48\linewidth}
  \centering
  \centerline{\includegraphics[height=3.3cm]{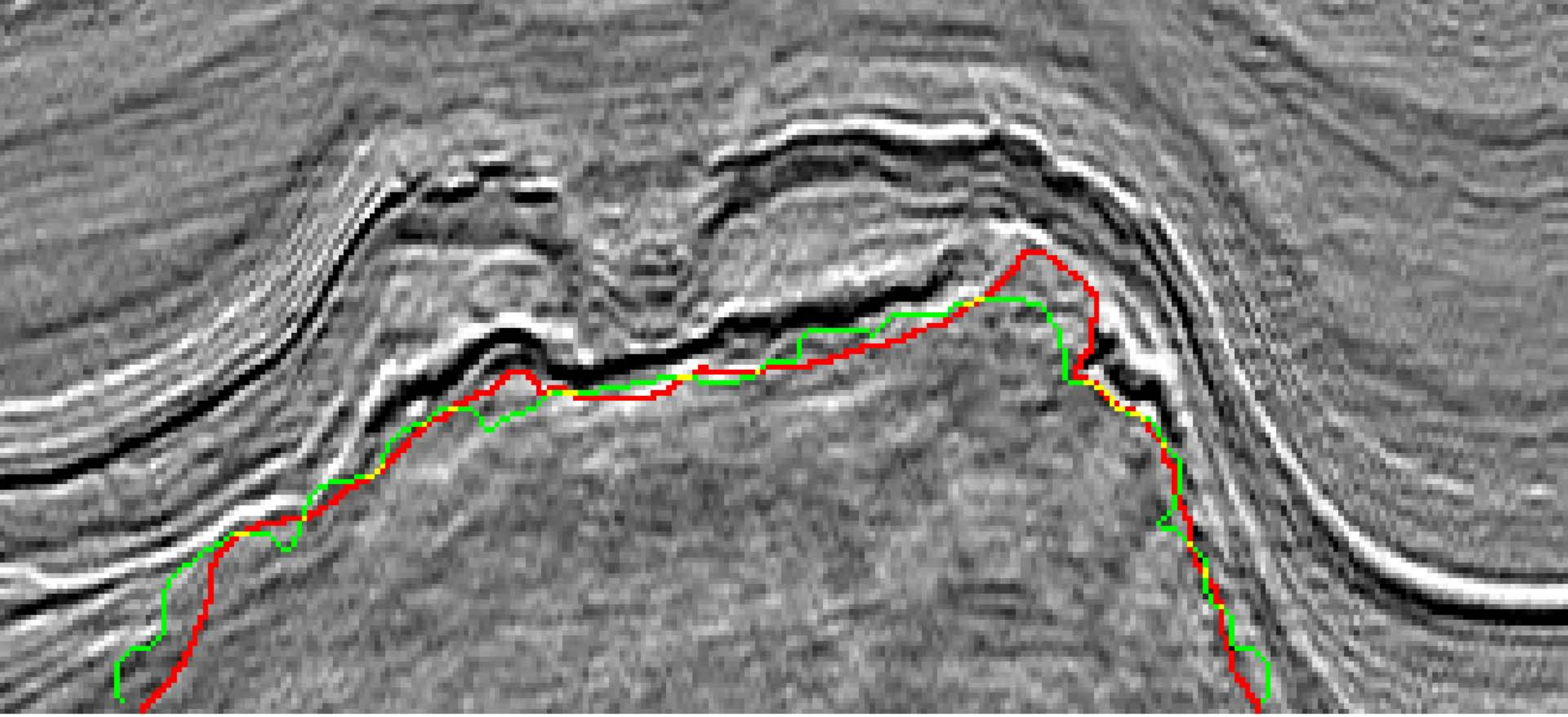}}
  \centerline{(d)}\medskip
\end{minipage}
\begin{minipage}[b]{0.5\linewidth}
  \centering
  \centerline{\includegraphics[height=3.3cm]{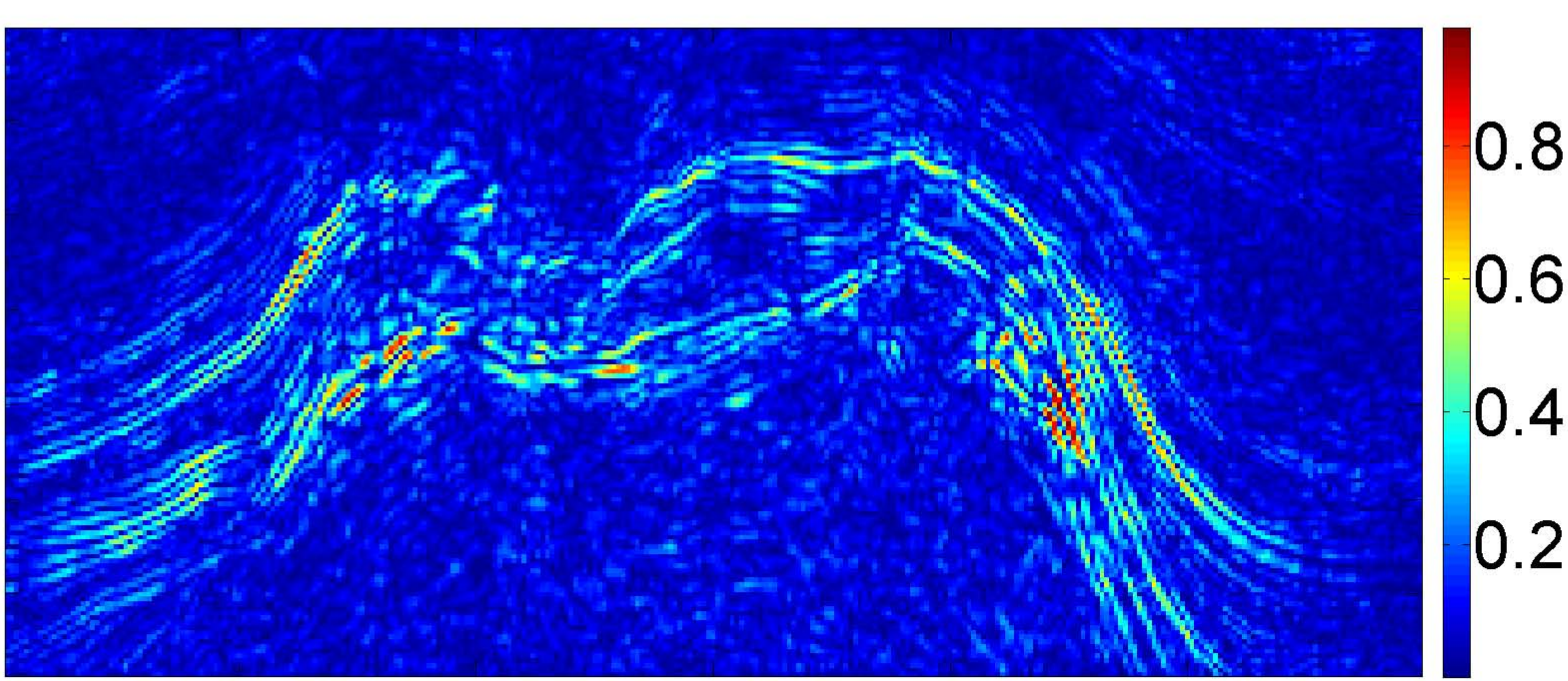}}
  \centerline{(e)}\medskip
\end{minipage}
\begin{minipage}[b]{0.48\linewidth}
  \centering
  \centerline{\includegraphics[height=3.3cm]{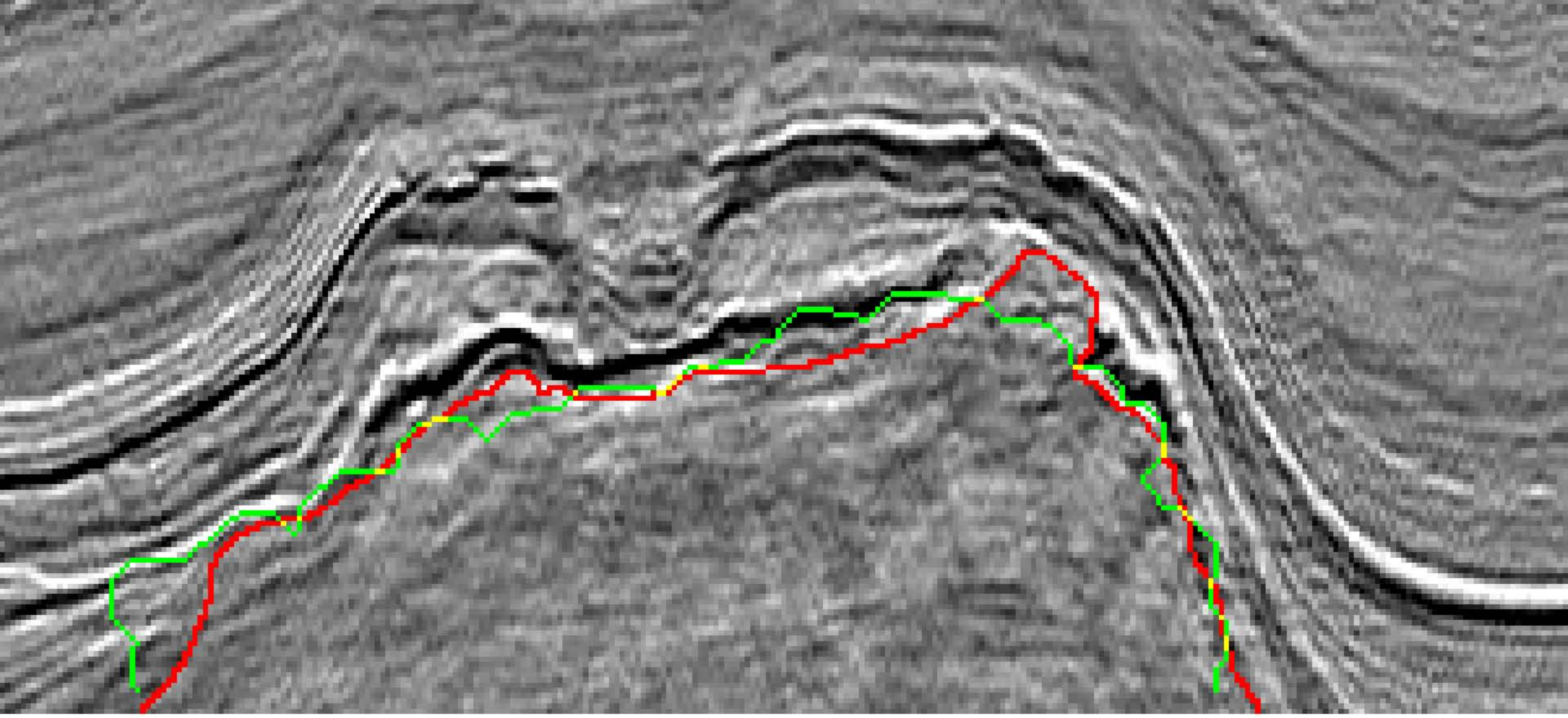}}
  \centerline{(f)}\medskip
\end{minipage}
\caption{(a), (c), and (e) illustrate the GoT, GLCM contrast, and gradient maps of Inline \#400, respectively; (b), (d), and (f) compare the manually labeled red ground truth with the green salt dome boundaries detected from attribute maps, (a), (c), and (e), respectively.}
\label{fig:detection}
\end{figure*}

Since GoT maps mainly characterize the texture variations of local neighborhoods in seismic sections, to demonstrate the benefit of the GoT attribute, we compare it with two conventional seismic attributes, the GLCM contrast attribute~\cite[]{haralick1973textural} and the gradient attribute~\cite[]{toriwaki2009fundamentals}, both of which have been introduced to detect salt dome boundaries \cite[]{berthelot2013texture,aqrawi2011detecting}. To ensure fair comparison, we apply the same region-growing method to extract salt dome boundaries from different attribute maps. The GLCM describes the distribution of co-occurring grayscale values at a given offset over an image. For each point $[i,j]$, within its $(2R_d+1)\times (2R_d+1)$ analysis window, by varying the directions and pixel distances of offset $[\Delta i, \Delta j]$, we can obtain a series of GLCMs, denoted $\mathbf{P_{[\Delta i, \Delta j]}}$. Therefore, on the basis of these GLCMs, we define the GLCM contrast attribute $\overline{\mathbf{C}}[i,j]$ as
\begin{equation}
\label{equ:glcmcont}
\Scale[0.97]{
\overline{\mathbf{C}}[i,j] = \sum\limits_{[\Delta i, \Delta j]}\left(\frac{1}{N_g}\sum\limits_{u}\sum\limits_{v}(u-v)^2\mathbf{P_{[\Delta i, \Delta j]}}[u,v]\right)},
\end{equation}
where $u$ and $v$ indicate the row and column indices of the GLCM, respectively, and $N_g$ represents the number of all possible offsets. In the tested local volume, we obtain the GLCM contrast maps of seismic sections by setting $R_d=4$. Figure~\ref{fig:detection}(c) illustrates the GLCM contrast attribute of Inline \#400, in which higher contrast values indicate likely boundary areas. By applying the region-growing method to Figure~\ref{fig:detection}(c), we can label the salt dome boundary as Figure~\ref{fig:detection}(d) shows. Similar to the GLCM contrast attribute, the gradient attribute estimated based on 3D Sobel filter can also identify the boundaries of salt domes. To approximate the partial derivatives of different directions, we convolute three $3\times 3\times 3$ kernels of the Sobel filter with the local volume. For each point, the magnitude of the gradient attribute is calculated as follows:
\begin{equation}
\label{equ:3dsobel}
F=\sqrt{F_x^2+F_y^2+F_z^2},
\end{equation}
where $F_x$, $F_y$, and $F_z$ represent the partial derivatives of the crossline, depth, and inline directions, respectively. Figure~\ref{fig:detection}(e) illustrates the gradient map of Inline \#400, from which we extract the green salt dome boundary as Figure~\ref{fig:detection}(f) depicts.

\begin{figure}[b]
\centering
\begin{minipage}[b]{\linewidth}
  \centering
  \centerline{\includegraphics[height=1.95cm]{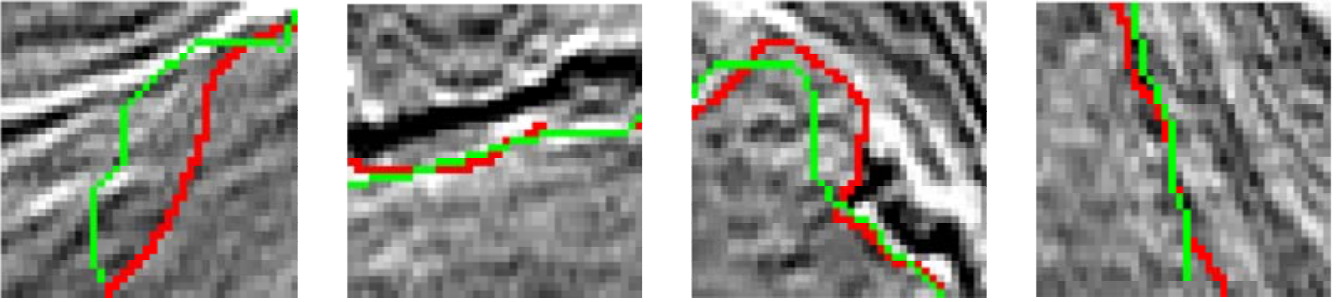}}
  \centerline{(a) Local boundaries extracted from the GoT map}\medskip
\end{minipage}
\begin{minipage}[b]{\linewidth}
  \centering
  \centerline{\includegraphics[height=1.95cm]{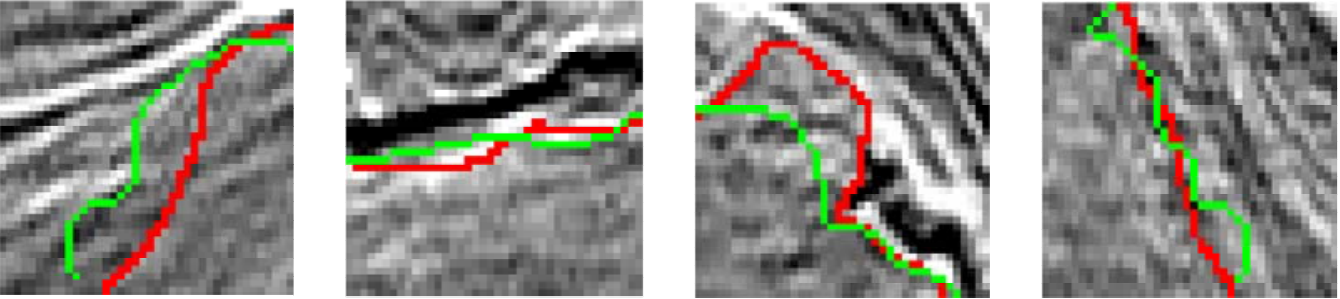}}
  \centerline{(b) Local boundaries extracted from the GLCM contrast map}\medskip
\end{minipage}
\begin{minipage}[b]{\linewidth}
  \centering
  \centerline{\includegraphics[height=1.95cm]{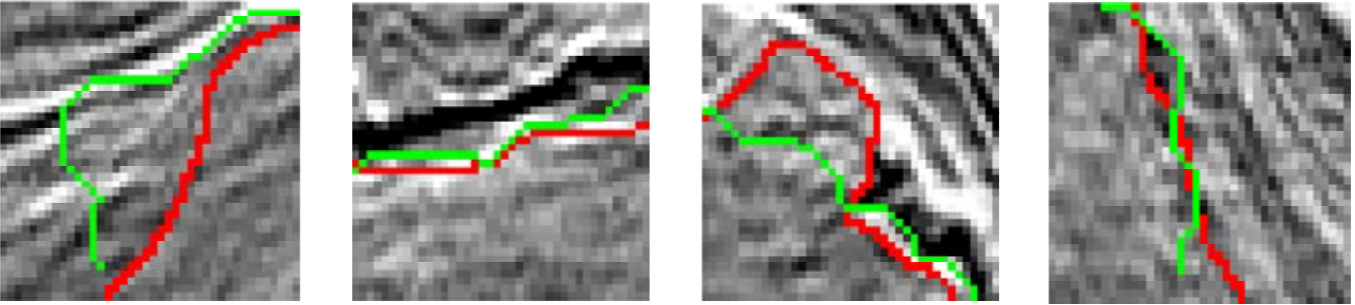}}
  \centerline{(c) Local boundaries extracted from the gradient map}\medskip
\end{minipage}
\caption{Local salt dome boundaries extracted from the GoT, GLCM contrast, and gradient maps are labeled in green. The ground truth labeled manually is labeled in red.}
\label{fig:localResult}
\end{figure}

By comparing the GoT and GLCM contrast of Inline \#400 in Figure~\ref{fig:detection}, we notice that it is not easy to determine which one leads to the more accurate detection of salt dome boundaries because of their comparable performance on highlighting boundary areas.
However, without multiscale-based description, the gradient map depending only on $3\times 3$ local regions shows noisy and discontinuous stripes around boundary areas, which limits the accuracy of detected boundaries. Figure~\ref{fig:localResult} illustrates the local regions of Figures~\ref{fig:detection}(b), (d), and (f), in which every column contains the same local regions of salt dome boundaries detected from various attribute maps and every row contain the different local regions of a detected salt dome boundary. The first column illustrates the bottom-left corner of detected boundaries, in which the second one is slightly more similar to the ground truth than the other two local boundaries. However, in second to fourth columns, boundaries detected by the proposed method show the highest accuracy.


\subsubsection{Objective Comparison of Detected Boundaries}
\label{sssec:detection_objective}

To quantize subjective perception, we employ the SalSIM index to evaluate the similarity between the ground truth and the detected boundary. Figure~\ref{fig:detection_cmpr} shows the SalSIM indices of the salt dome boundaries of Inline \#389 to \#409 detected by various methods. 
Boundaries extracted from gradient maps have the lowest accuracy because of noise around boundary areas.
In most of the tested seismic sections, boundaries detected by the proposed method have SalSIM indices higher than or equal to those obtained from GLCM contrast maps. Since the computation of GLCMs is time consuming, the proposed method outperforms the GLCM-based detection method on both efficiency and accuracy. Table~\ref{tab:detection} lists several statistical measures of SalSIM indices, in which the averaged maximum distance (AMD) in pixels represents the mean of the $d_{max}$ of all detected boundaries. The SalSIM indices of boundaries detected by the proposed method has the largest mean, the smallest standard deviation, and the shortest AMD, which proves the superiority of the proposed method.

To further verify the robustness of the proposed method, we apply the GoT-based detection method on the ten sections of the F3 block with crossline number ranging from \#834 to \#842, inline number ranging from \#279 to \#600, and the time interval between 1,300ms and 1,848ms sampled every 4ms. Figure~\ref{fig:crossline} illustrates one section of this local dataset, and Table~\ref{tab:detection_cross} shows the performance of various detection methods, in which the proposed GoT-based method still achieves the highest accuracy.

\begin{table*}[!htbp]
\caption{The statistical measures of SalSIM indices in Figure~\ref{fig:detection_cmpr} obtained from various detection methods.}
\footnotesize
  \begin{center}
  \label{tab:detection}
    \begin{tabular}{|c|c|c|c|}
      \hline
      \multirow{2}{*}{Detection Methods} & \multirow{2}{*}{Mean} & Standard & \multirow{2}{*}{AMD (pixels)}\\
      & & Deviation &\\
      \hline
      Detection method & \multirow{2}{*}{\textbf{0.9348}}
       & \multirow{2}{*}{\textbf{0.0104}} & \multirow{2}{*}{\textbf{11.64}}\\
       based on GoT Maps & & & \\
      \hline
      Detection method based & \multirow{2}{*}{0.9290} & \multirow{2}{*}{\textbf{0.0104}} & \multirow{2}{*}{13.92}\\
      on GLCM Contrast Maps~\cite[]{berthelot2013texture} & & &\\
      \hline
      Detection method based on & \multirow{2}{*}{0.9054} & \multirow{2}{*}{0.0159} & \multirow{2}{*}{20.73}\\
      Gradient Maps~\cite[]{aqrawi2011detecting} & & &\\
      \hline
    \end{tabular}
  \end{center}
\end{table*}

\begin{table*}[!htbp]
\caption{The statistical measures of the SalSIM indices of boundaries delineated by various detection methods in the crossline dataset.}
\footnotesize
  \begin{center}
  \label{tab:detection_cross}
    \begin{tabular}{|c|c|c|c|}
      \hline
      \multirow{2}{*}{Detection Methods} & \multirow{2}{*}{Mean} & Standard & \multirow{2}{*}{AMD (pixels)}\\
      & & Deviation &\\
      \hline
      Detection method & \multirow{2}{*}{\textbf{0.9678}}
       & \multirow{2}{*}{\textbf{0.00211}} & \multirow{2}{*}{\textbf{4.16}}\\
       based on GoT Maps & & & \\
      \hline
      Detection method based & \multirow{2}{*}{0.9596} & \multirow{2}{*}{0.00326} & \multirow{2}{*}{6.57}\\
      on GLCM Contrast Maps~\cite[]{berthelot2013texture} & & &\\
      \hline
      Detection method based on & \multirow{2}{*}{0.9410} & \multirow{2}{*}{0.00822} & \multirow{2}{*}{8.74}\\
      Gradient Maps~\cite[]{aqrawi2011detecting} & & &\\
      \hline
    \end{tabular}
  \end{center}
\end{table*}

\subsubsection{Robustness to Noise}
\begin{figure*}[b]
\centering
\begin{minipage}[b]{\linewidth}
  \centering
  \centerline{\includegraphics[width=12cm]{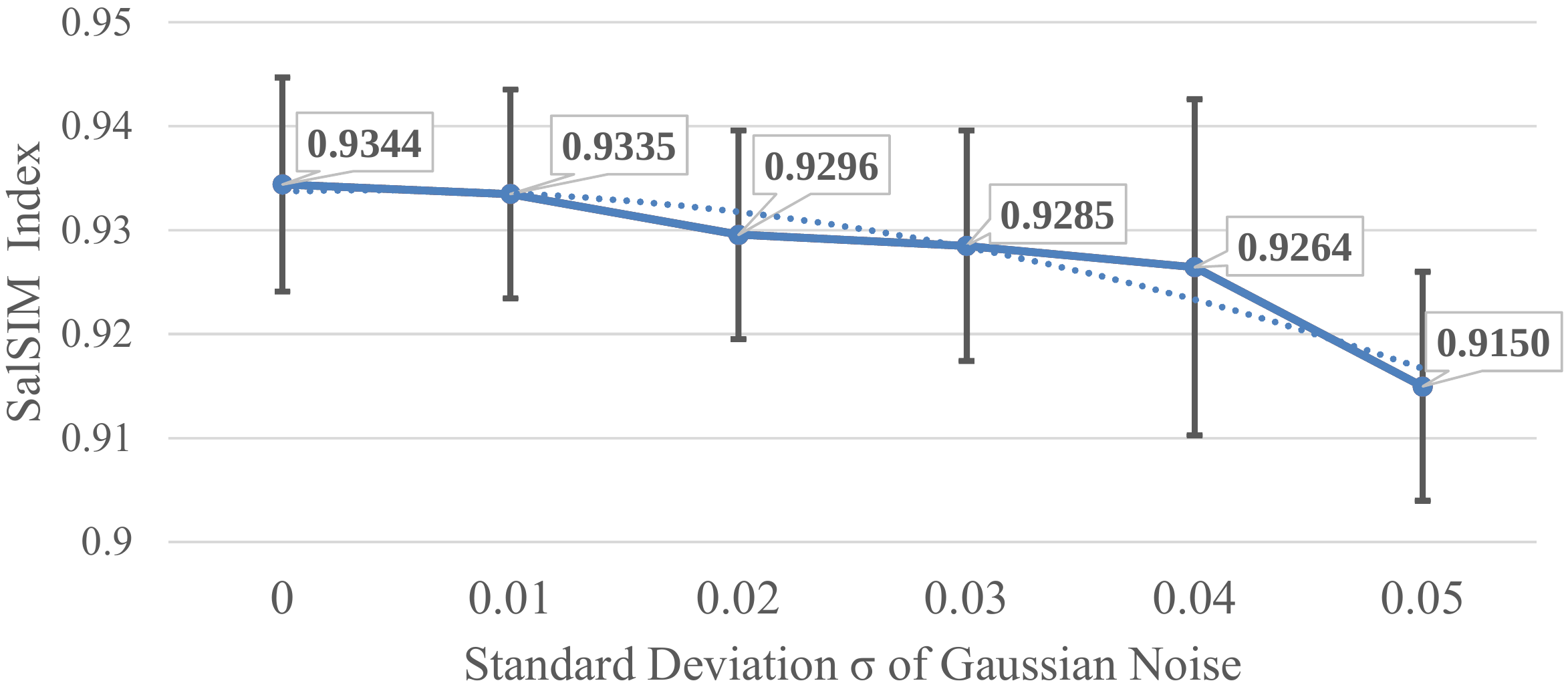}}
  \centerline{(a)The proposed method without denoising}\medskip
\end{minipage}
\begin{minipage}[b]{\linewidth}
  \centering
  \centerline{\includegraphics[width=12cm]{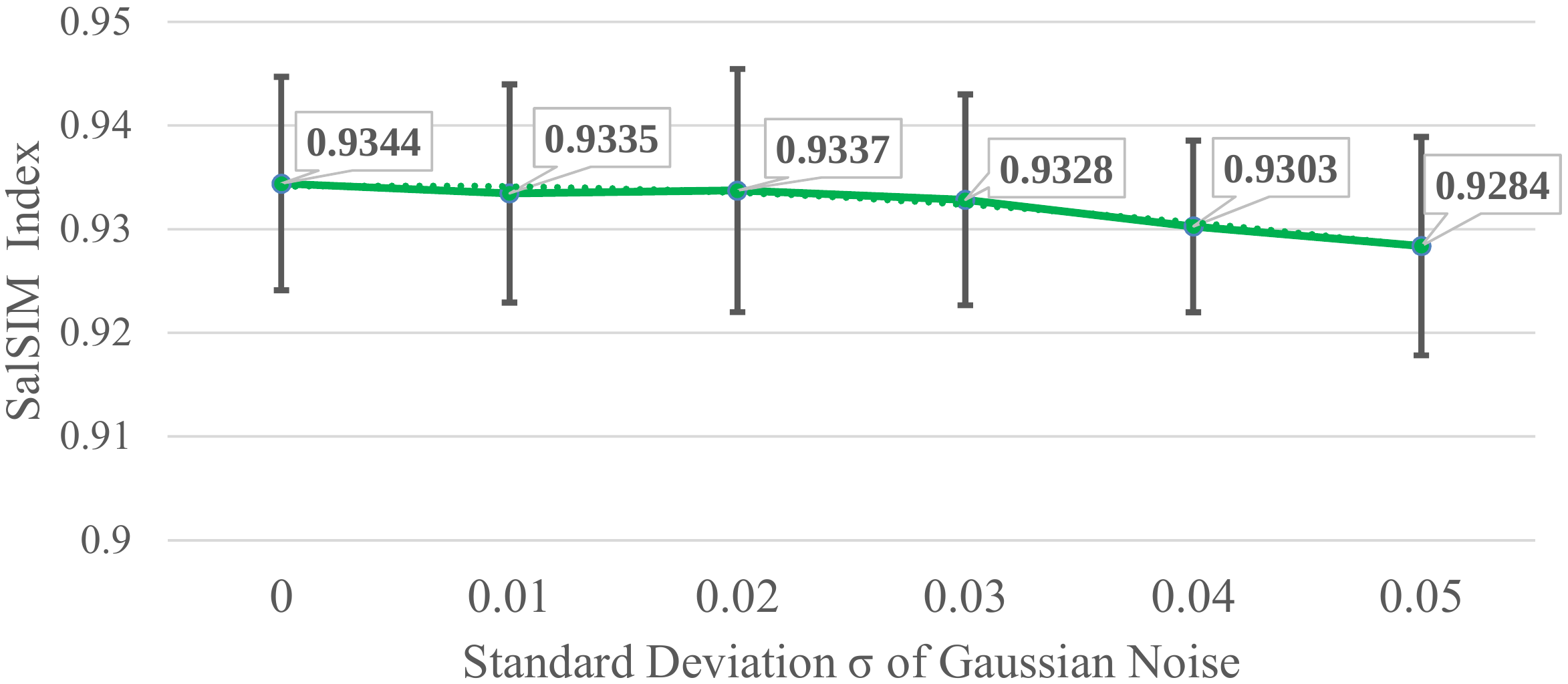}}
  \centerline{(b)The proposed method based on denoising operation}\medskip
\end{minipage}
\begin{minipage}[b]{\linewidth}
  \centering
  \centerline{\includegraphics[width=12cm]{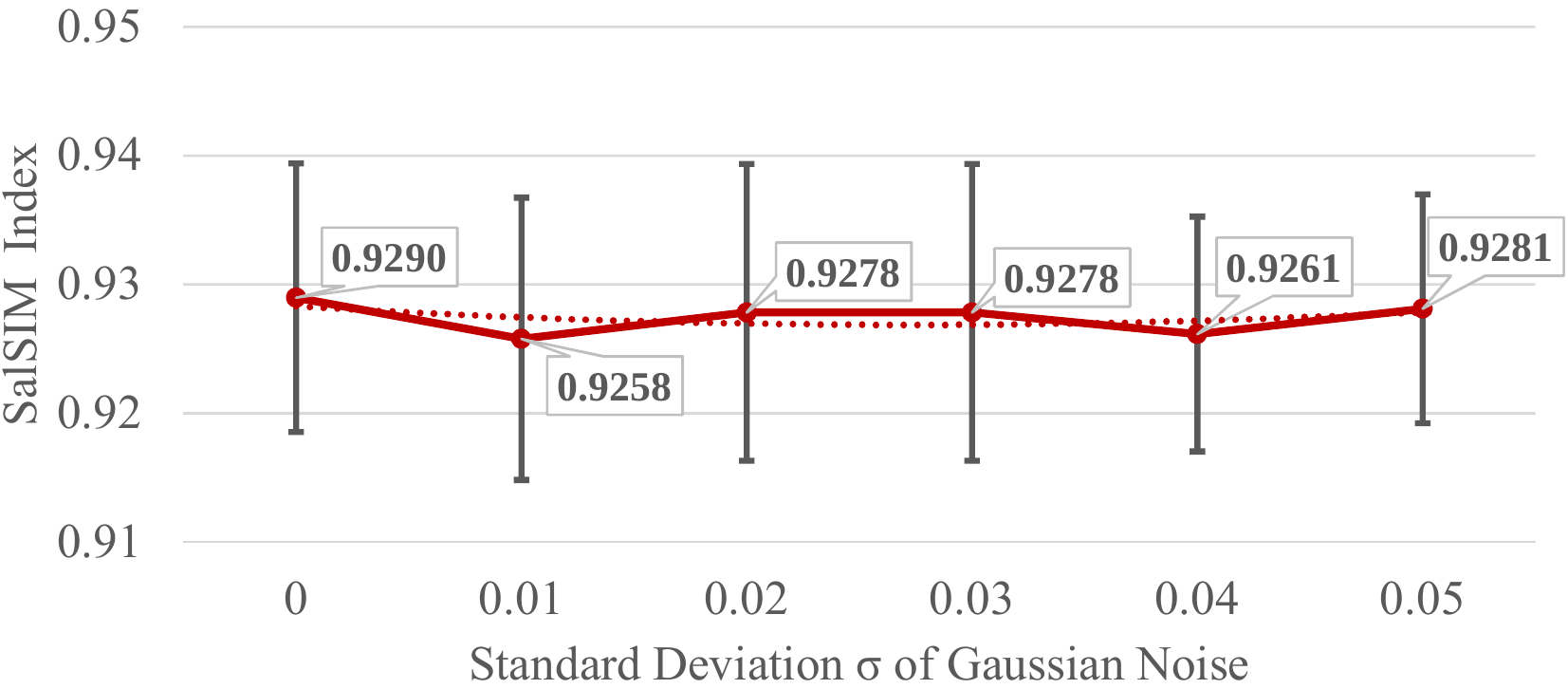}}
  \centerline{(c)The GLCM-based detection method}\medskip
\end{minipage}
\caption{The averaged SalSIM indices of salt dome boundaries detected from noisy seismic sections based on the proposed method and the GLCM-based method.}
\label{fig:detectionNoise}
\end{figure*}

To verify the robustness of the proposed method, we add Gaussian noise to seismic sections and compare the accuracy of salt dome boundaries detected from the corresponding GoT and GLCM contrast maps. The added zero-mean Gaussian noise has standard deviations ranging from $0.01$ to $0.05$, and Figure~\ref{fig:detectionNoise} illustrates the change of SalSIM indices according to the increasing of standard deviations. For one noise level, we extract the salt dome boundaries from noisy seismic sections using either the proposed method or the GLCM-based detection method, and the mean of corresponding SalSIM indices is shown in Figure~\ref{fig:detectionNoise}. The error bars of means correspond to an uncertainty equal to one standard deviation, and the dashed curve represents the trend of means. Although the averaged accuracy of salt dome boundaries detected by the proposed method decreases slowly with the increasing of standard deviations, it is still greater than those of boundaries detected by the GLCM-based method at the most of noisy levels. We notice that with the increasing of noise the latter fluctuates around a certain SalSIM index rather than keep decreasing, the reason for which is that the quantization step in the calculation of the GLCM, functioning as a built-in noise filter, can weaken the influence of noise on the GLCM contrast maps. To fairly compare the robustness of these two methods, we apply an edge-preserving smoothing filter, the bilateral filter~\cite[]{tomasi1998bilateral}, to noisy seismic sections and obtain the statistical measures of corresponding SalSIM indices in Figure~\ref{fig:detectionNoise}(a). We notice that with the extra denoising operation, the GoT-based method shows higher accuracy than the GLCM-based method. If the noise level keeps increasing, depending only on the denoising operation we may not be able to label satisfied boundaries from migrated seismic sections. Therefore, geophysicists commonly prefer to regenerate migrated seismic sections with higher quality from raw seismic data by reducing noise that appears in every preprocessing step~\cite[]{chopra2014noise}.

\subsection{Salt Dome Tracking}
In the local seismic volume we define Inline \#400 as a reference section, the salt dome boundary of which has been labeled by experienced interpreters or computer-aided detection methods. We can track the reference boundaries through the seismic volume and synthesize salt dome boundaries in the neighboring twenty predicted sections ranging from Inline \#389 to \#409. Each point at the reference boundary corresponds to a pair of $31\times31$ image patches extracted from the reference section and its corresponding GoT map. To acquire texture features from the local regions of the reference boundary, we group all these patch pairs into tensor pairs on the basis of the block diagram shown in Figure~\ref{fig:classification}. The dimensions of texture features in three modes are $[15,15,5]$, and threshold $T_e$ on the reconstruction error is $2.3$. Furthermore, in predicted sections the proposed tracking method searches along the normal direction of the projected point and identifies the position of the tracked point by comparing the similarity between the patch pairs of candidate points and tensor pairs built from the reference section. The localization of tracked points is implemented automatically, which improves interpretation efficiency. Finally, we remove noisy points in predicted sections with $2\times 2$ median filters and connect remaining points to label the salt dome boundary under the shape constraint of salt domes. Figure~\ref{fig:tracking}(a) compares the green tracked salt dome boundary in Inline \#391, synthesized based on the manually labeled reference boundary in Inline \#400, with the red ground truth. We notice that these two curves almost overlap except for several local regions.
%
%
\subsubsection{Subjective Comparison of Tracking Methods}
\label{sssec:tracking_subjective}

\begin{figure*}[b]
\centering
\begin{minipage}[b]{0.5\linewidth}
  \centering
  \centerline{\includegraphics[height=3.3cm]{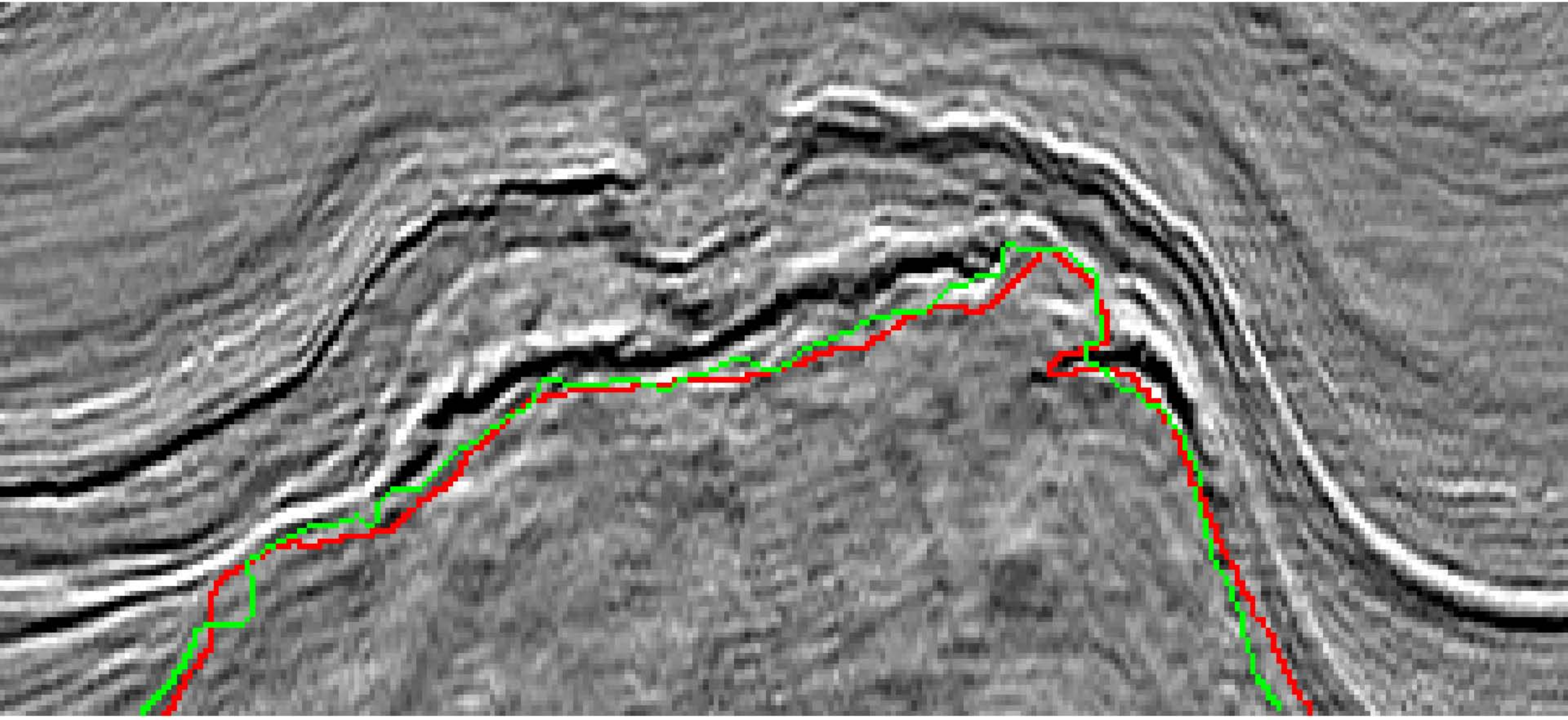}}
  \centerline{\scriptsize{(a) proposed tracking method}}\medskip
\end{minipage}
\begin{minipage}[b]{0.48\linewidth}
  \centering
  \centerline{\includegraphics[height=3.3cm]{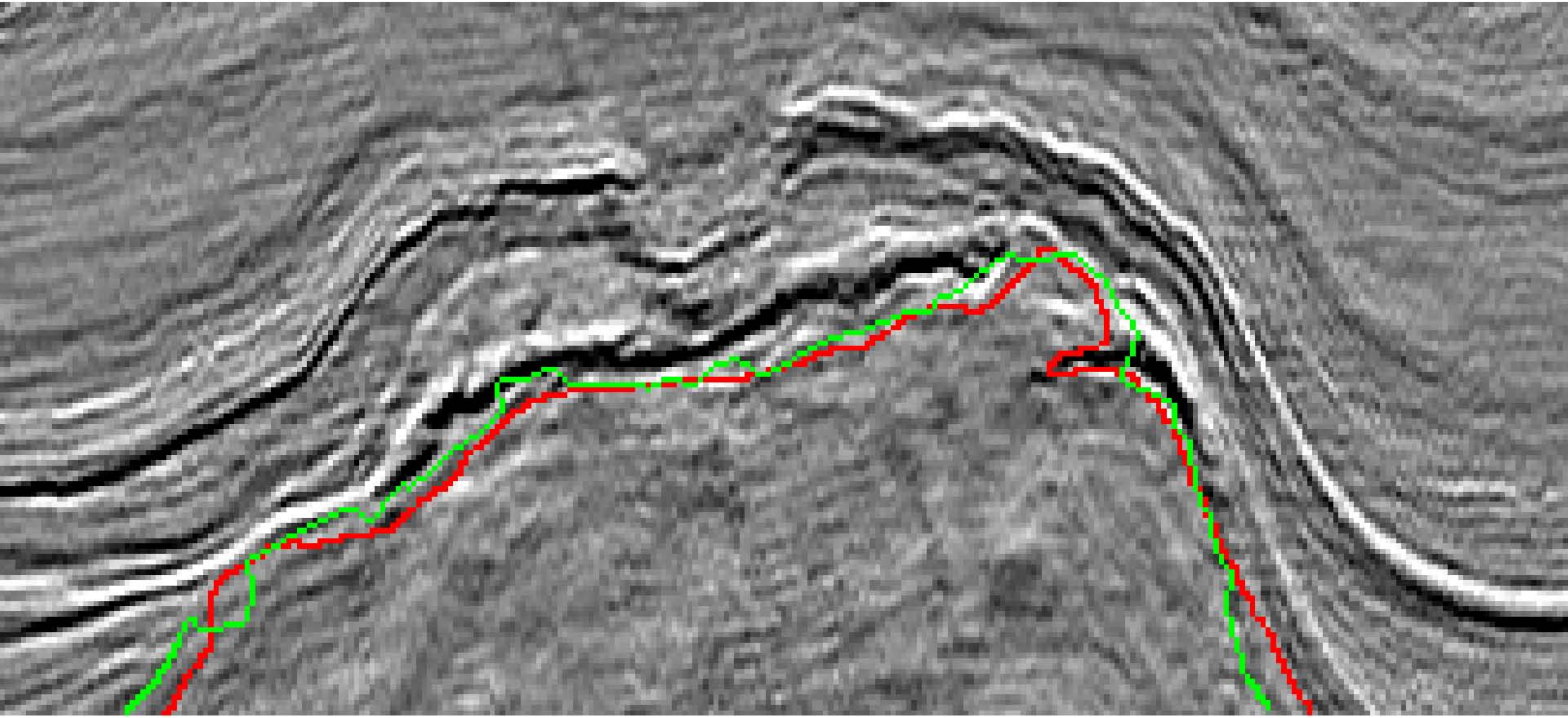}}
  \centerline{\scriptsize{(b) tracking method based on vectorization}}\medskip
\end{minipage}
\begin{minipage}[b]{0.5\linewidth}
  \centering
  \centerline{\includegraphics[height=3.3cm]{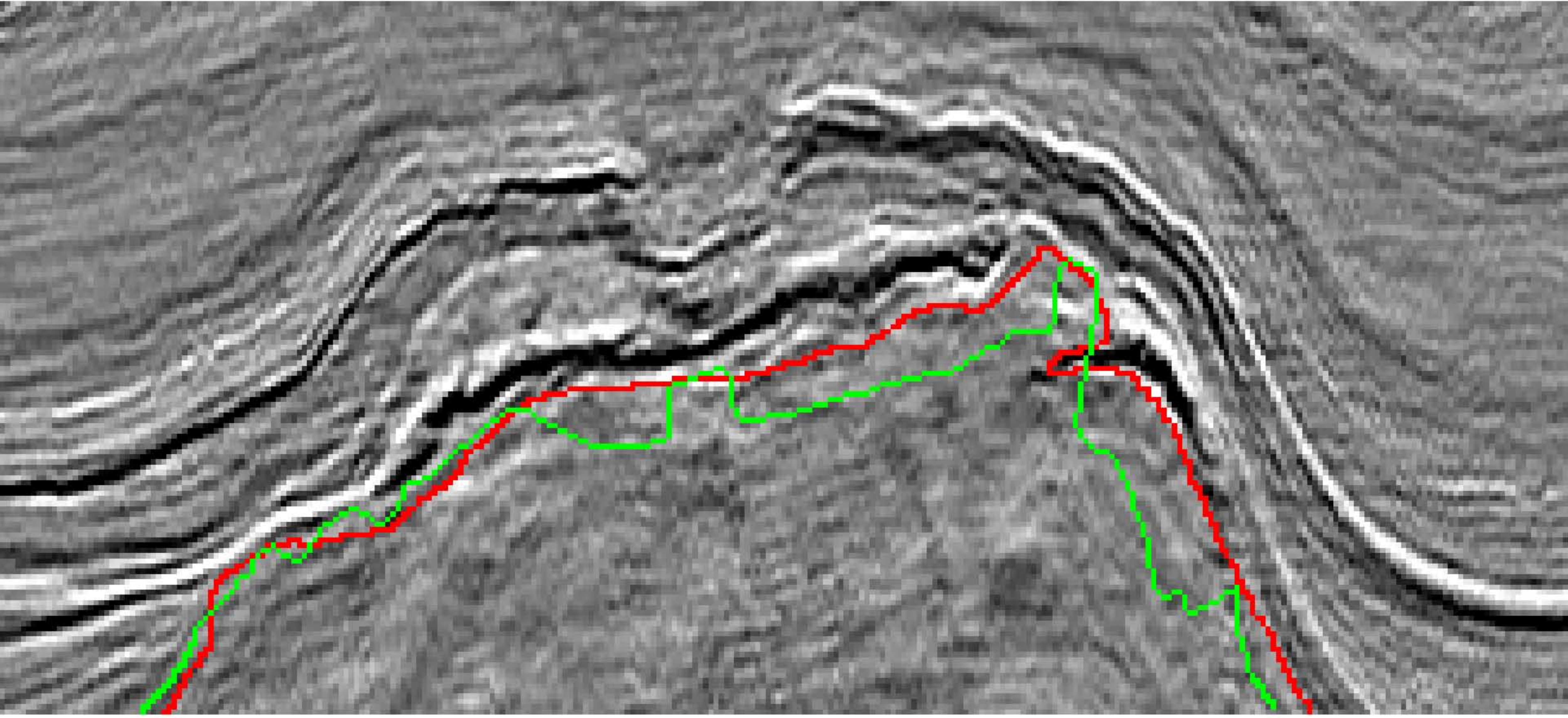}}
  \centerline{\scriptsize{(c) tensor-based tracking method without GoT maps}}\medskip
\end{minipage}
\begin{minipage}[b]{0.48\linewidth}
  \centering
  \centerline{\includegraphics[height=3.3cm]{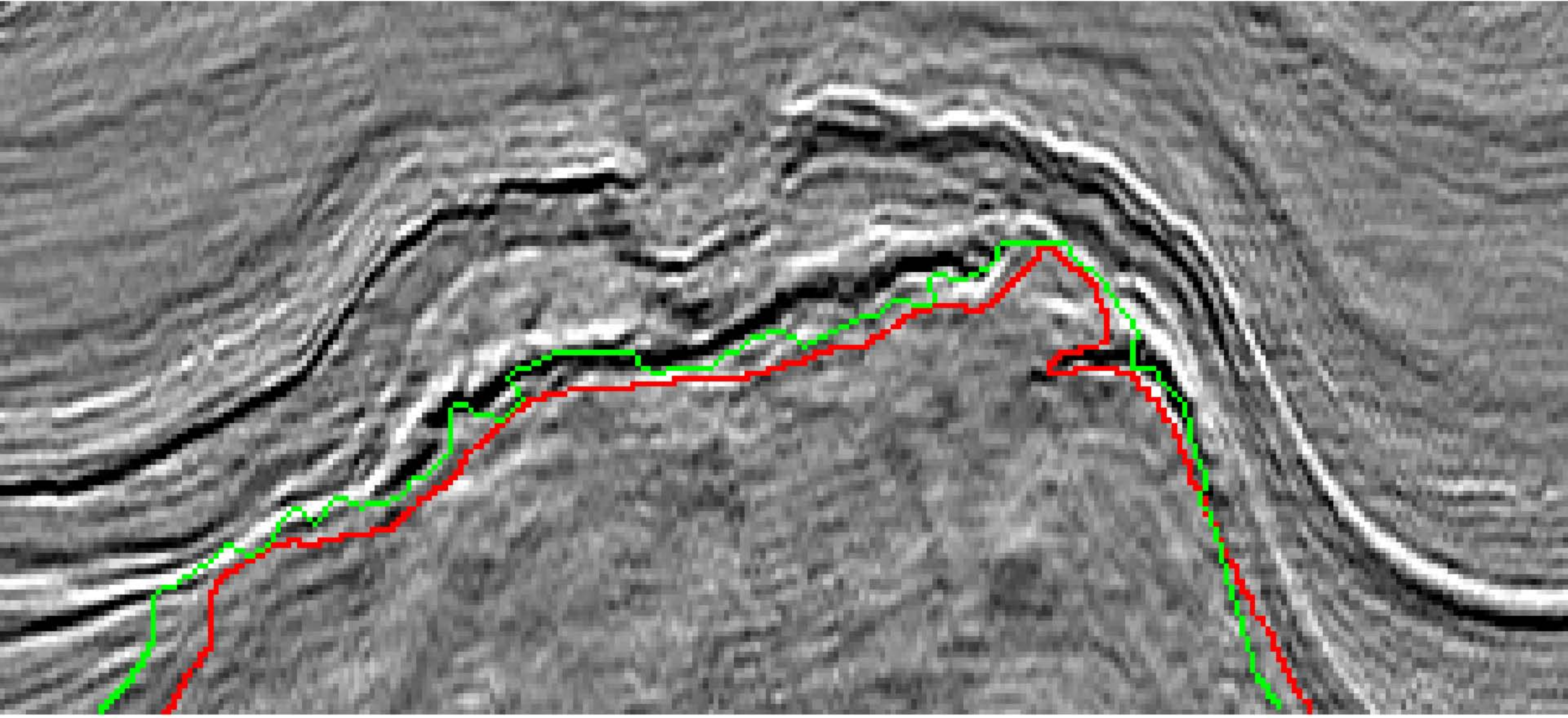}}
  \centerline{\scriptsize{(d) tensor-based tracking method with GLCM contrast maps}}\medskip
\end{minipage}
\caption{(a), (b), (c), and (d) show the comparison between the red ground truth and the green tracked salt dome boundaries of Inline \#391 synthesized by the proposed method, the tracking method based on vectorization, the tensor-based tracking method without GoT maps, and the tensor-based tracking method with GLCM contrast maps ,  respectively. }
\label{fig:tracking}
\end{figure*}

\begin{figure}[b]
\centering
\begin{minipage}[b]{\linewidth}
  \centering
  \centerline{\includegraphics[height=1.95cm]{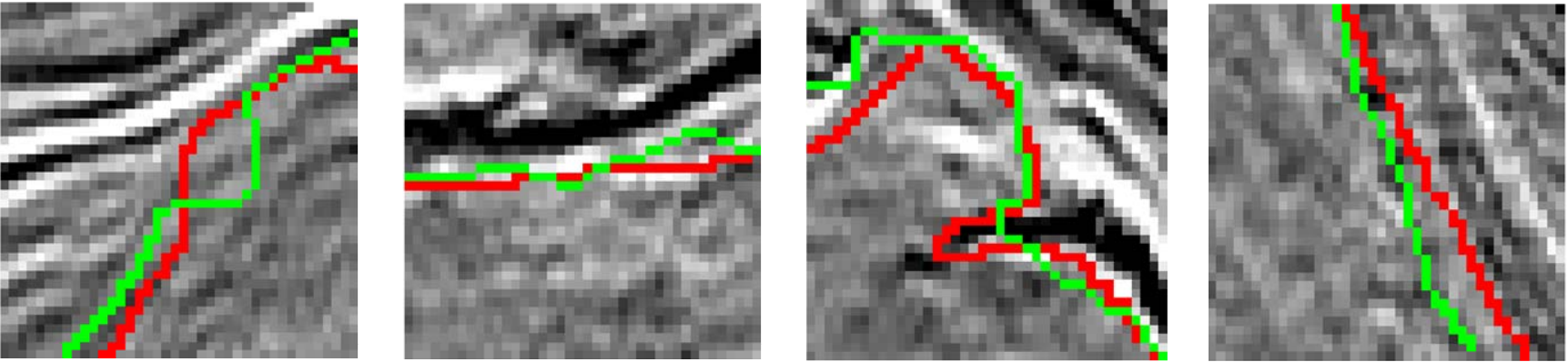}}
  \centerline{(a)}\medskip
\end{minipage}
\begin{minipage}[b]{\linewidth}
  \centering
  \centerline{\includegraphics[height=1.95cm]{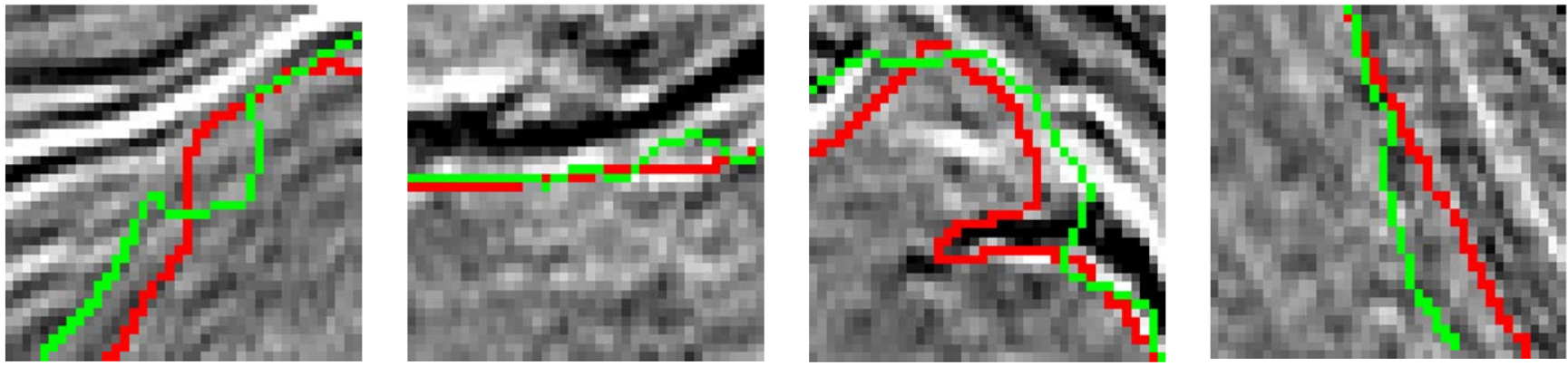}}
  \centerline{(b)}\medskip
\end{minipage}
\begin{minipage}[b]{\linewidth}
  \centering
  \centerline{\includegraphics[height=1.95cm]{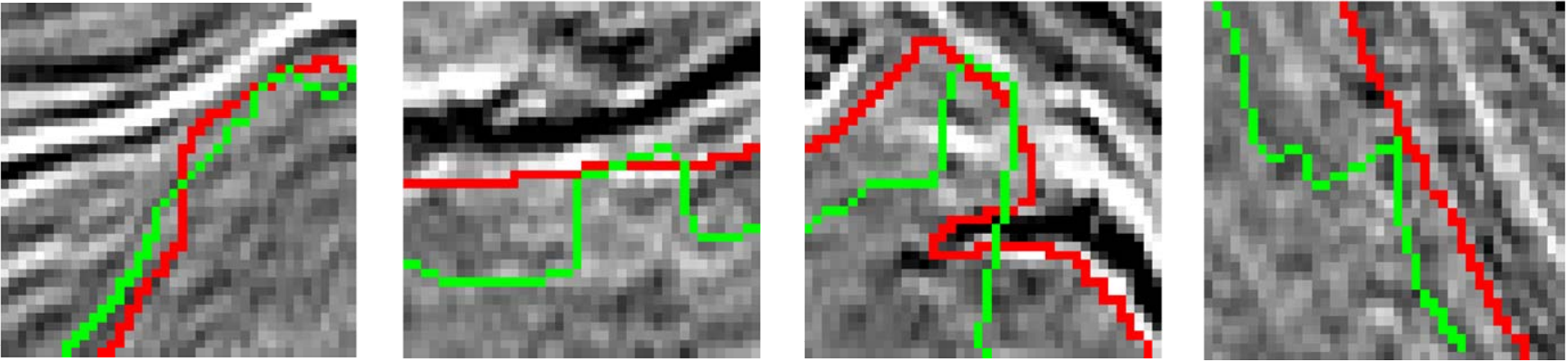}}
  \centerline{(c)}\medskip
\end{minipage}
\begin{minipage}[b]{\linewidth}
  \centering
  \centerline{\includegraphics[height=1.95cm]{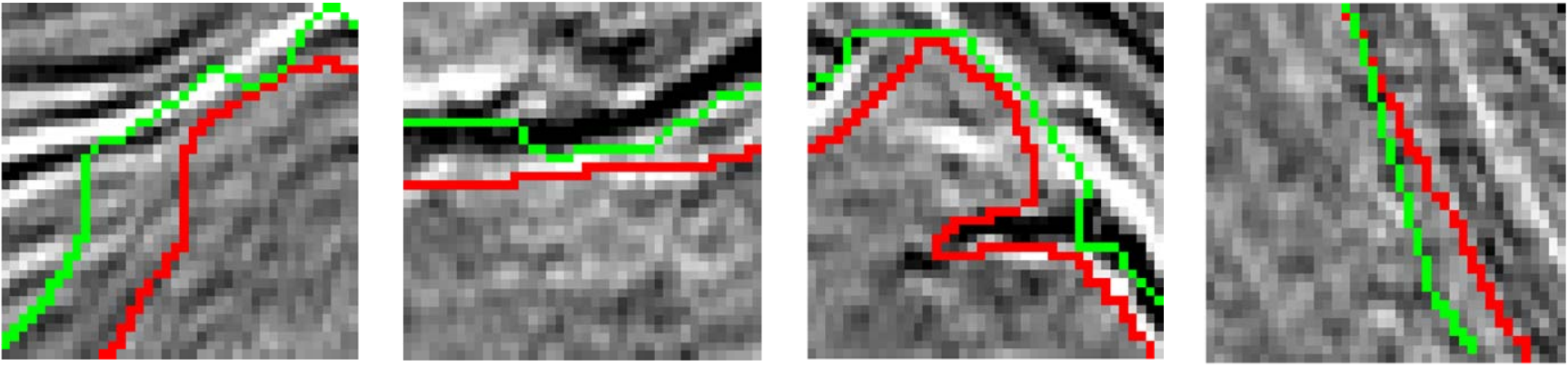}}
  \centerline{(d)}\medskip
\end{minipage}
\caption{(a), (b), (c), and (d) contain four local regions extracted from Figures~\ref{fig:tracking}(a) to (d), respectively. The red and green curves represent tracked salt dome boundaries and the ground truth, respectively.}
\label{fig:trackingLocal}
\end{figure}

To prove that tracking accuracy can be increased by involving the GoT attribute and the texture features of all modes, we need to compare the proposed method with three other tracking strategies. The first one is to implement the tracking process depending only on texture features extracted from vectorized patches rather than those from third-order tensors. The second one is to synthesize tracked boundaries within the framework of the proposed method, but without involving the GoT attribute. The third strategy is the same as the proposed  method except for using GLCM contrast maps instead of GoT maps. We rename these three tracking strategies as the tracking method based on vectorization, the tensor-based tracking method without GoT maps, and the tensor-based tracking method with GLCM contrast maps. Figures~\ref{fig:tracking}(b) to (d) compare the red ground truth labeled manually with green salt dome boundaries labeled by the three tracking strategies mentioned above, and the tracked boundary in Figure~\ref{fig:tracking}(c) shows the greatest deviation from the ground truth. Figure~\ref{fig:trackingLocal} illustrates the local regions of Figures~\ref{fig:tracking}(a) to (d), in which every column contains the same local regions of salt dome boundaries synthesized based on various tracking methods and every row contains the different local regions of a tracked salt dome boundary. By comparing the local regions of tracked boundaries with the ground truth, we notice the boundary in the first row synthesized by the proposed tracking method is the most similar to the ground truth.

\subsubsection{Objective Comparison of Tracking Methods}
\label{sssec:tracking_objective}
To objectively verify our conclusion, we synthesize tracked boundaries in seismic sections ranging from Inline \#389 to \#409 using various tracking strategies and plot their SalSIM indices in Figure~\ref{fig:tracking_cmpr}. The horizontal and vertical axes represent the inline number and the SalSIM index, respectively. Although tracking method based on vectorization can achieve comparable accuracy with the proposed method in Inline \#401 to \#409, the SalSIM indices of salt dome boundaries in Inline \#395 to \#389 synthesized by the former one decrease more quickly than those obtained by the latter one. In addition, dashed curves in Figure~\ref{fig:tracking_cmpr} shows the distribution trends of SalSIM indices. Although the SalSIM indices of four tracking strategies have similar trends that drop with the increasing offsets between predicted sections and the reference section, the proposed tracking method outperforms other tracking strategies particularly in Inline \#389 to \#399. Table~\ref{tab:tracking} contains several statistical measures of SalSIM indices in Figure~\ref{fig:tracking_cmpr}. The mean of SalSIM indices and the AMD evaluate the accuracy of the tracking method, and the standard deviation determines robustness. The proposed method has the greatest mean, the smallest standard deviation, and the shortest AMD, which proves the superiority of the proposed method.

\begin{table*}[!htbp]
\caption{The statistical measures of SalSIM indices in Figure~\ref{fig:tracking_cmpr} obtained from various tracking strategies.}
\footnotesize
  \begin{center}
  \label{tab:tracking}
    \begin{tabular}{|c|c|c|c|}
      \hline
      \multirow{2}{*}{Tracking Methods} & \multirow{2}{*}{Mean} & Standard & AMD\\
      & & Deviation &  (pixels)\\
      \hline
      \multirow{2}{*}{Proposed tracking method} & \multirow{2}{*}{\textbf{0.9531}}
       & \multirow{2}{*}{\textbf{0.0111}} & \multirow{2}{*}{\textbf{8.32}}\\
       & & &\\
      \hline
      Tracking method based & \multirow{2}{*}{0.9496} & \multirow{2}{*}{0.0148} & \multirow{2}{*}{9.30}\\
      on vectorization & & &\\
      \hline
      Tensor-based tracking method & \multirow{2}{*}{0.9364} & \multirow{2}{*}{0.0295} & \multirow{2}{*}{11.26}\\
      without GoT maps & & &\\
      \hline
      Tensor-based tracking method & \multirow{2}{*}{0.9422} & \multirow{2}{*}{0.0189} & \multirow{2}{*}{10.53}\\
      with GLCM contrast maps & & &\\
      \hline
    \end{tabular}
  \end{center}
\end{table*}

%


\subsubsection{Combination of Proposed Detection and Tracking Methods}
\label{sssec:combined}

\begin{figure*}[b]
\centering
\begin{minipage}[b]{0.5\linewidth}
  \centering
  \centerline{\includegraphics[height=3.3cm]{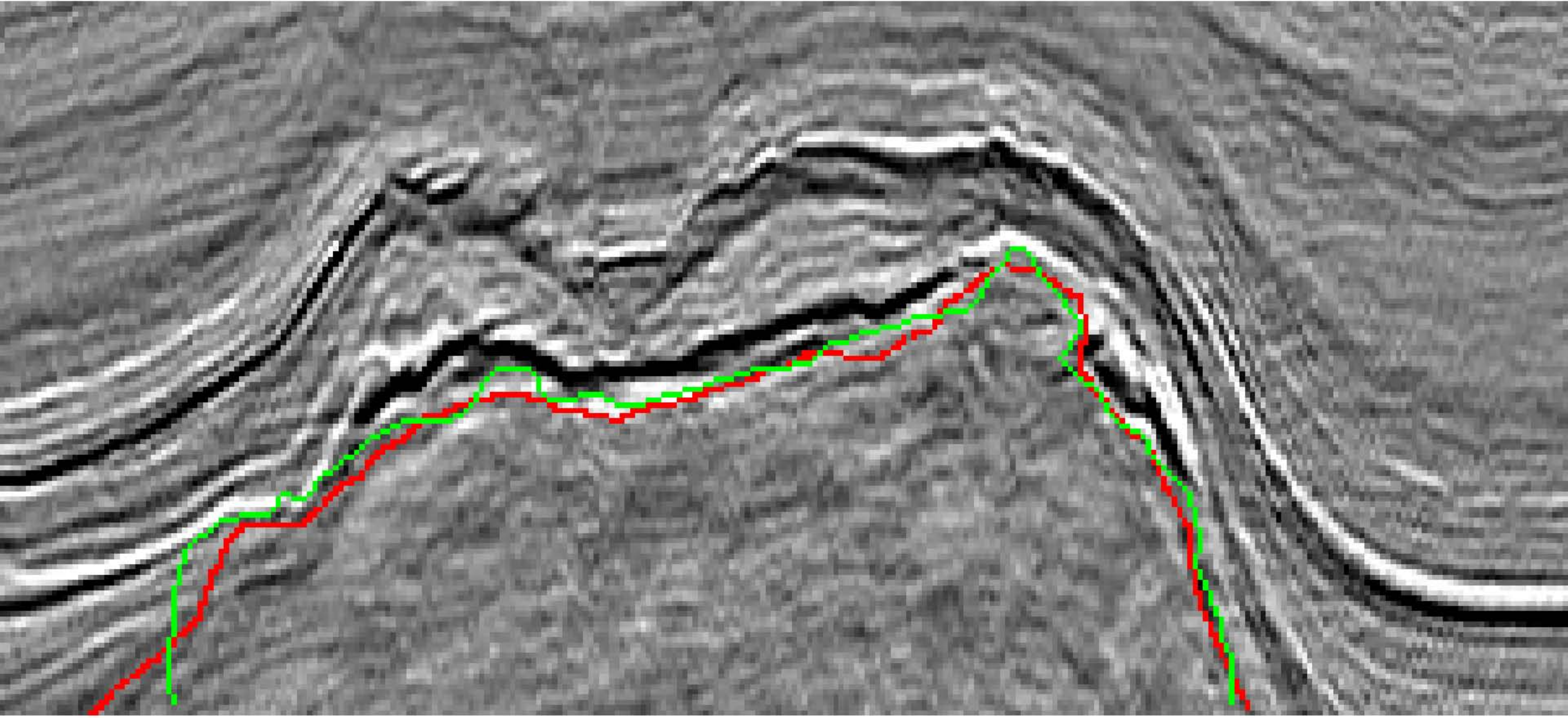}}
  \centerline{(a)}\medskip
\end{minipage}
\begin{minipage}[b]{0.48\linewidth}
  \centering
  \centerline{\includegraphics[height=3.3cm]{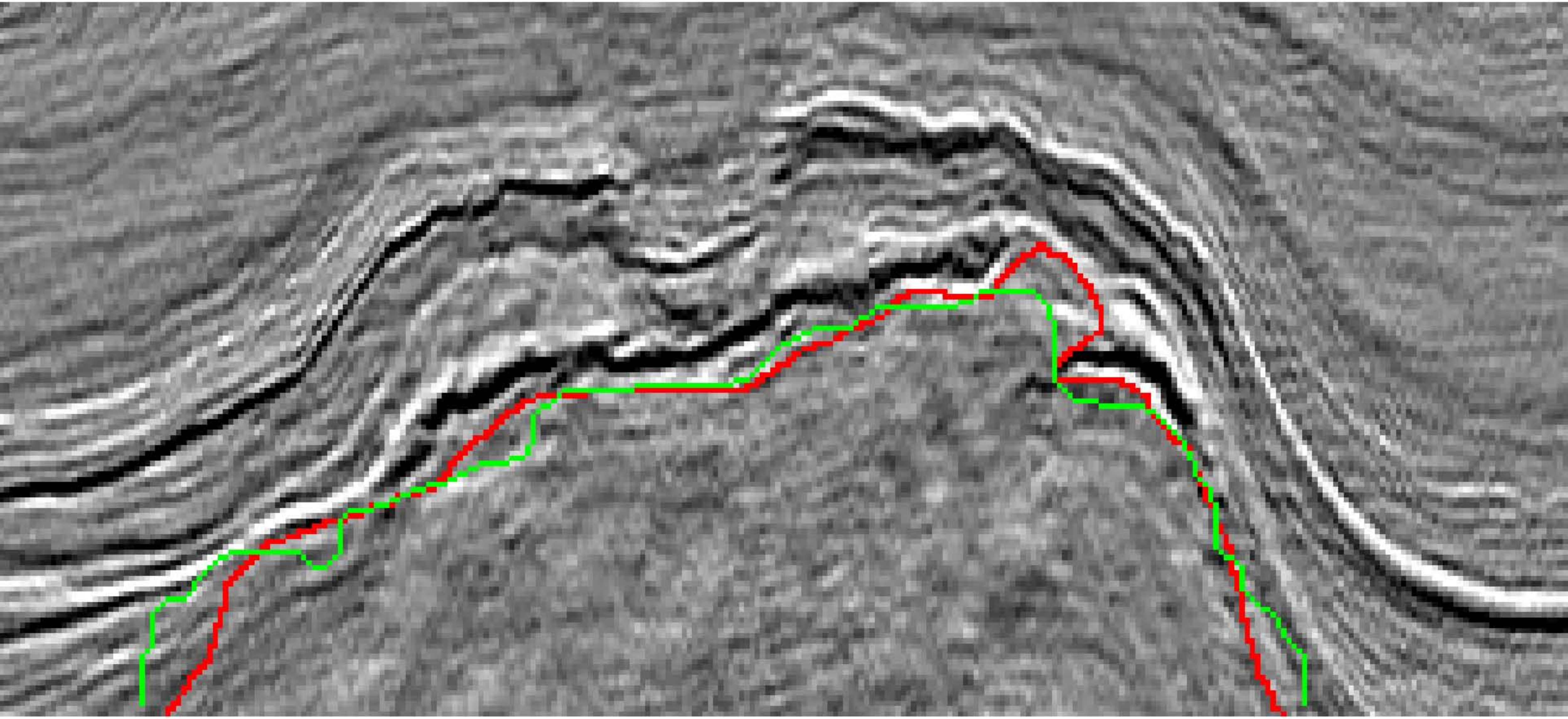}}
  \centerline{(b)}\medskip
\end{minipage}
\begin{minipage}[b]{0.5\linewidth}
  \centering
  \centerline{\includegraphics[height=3.3cm]{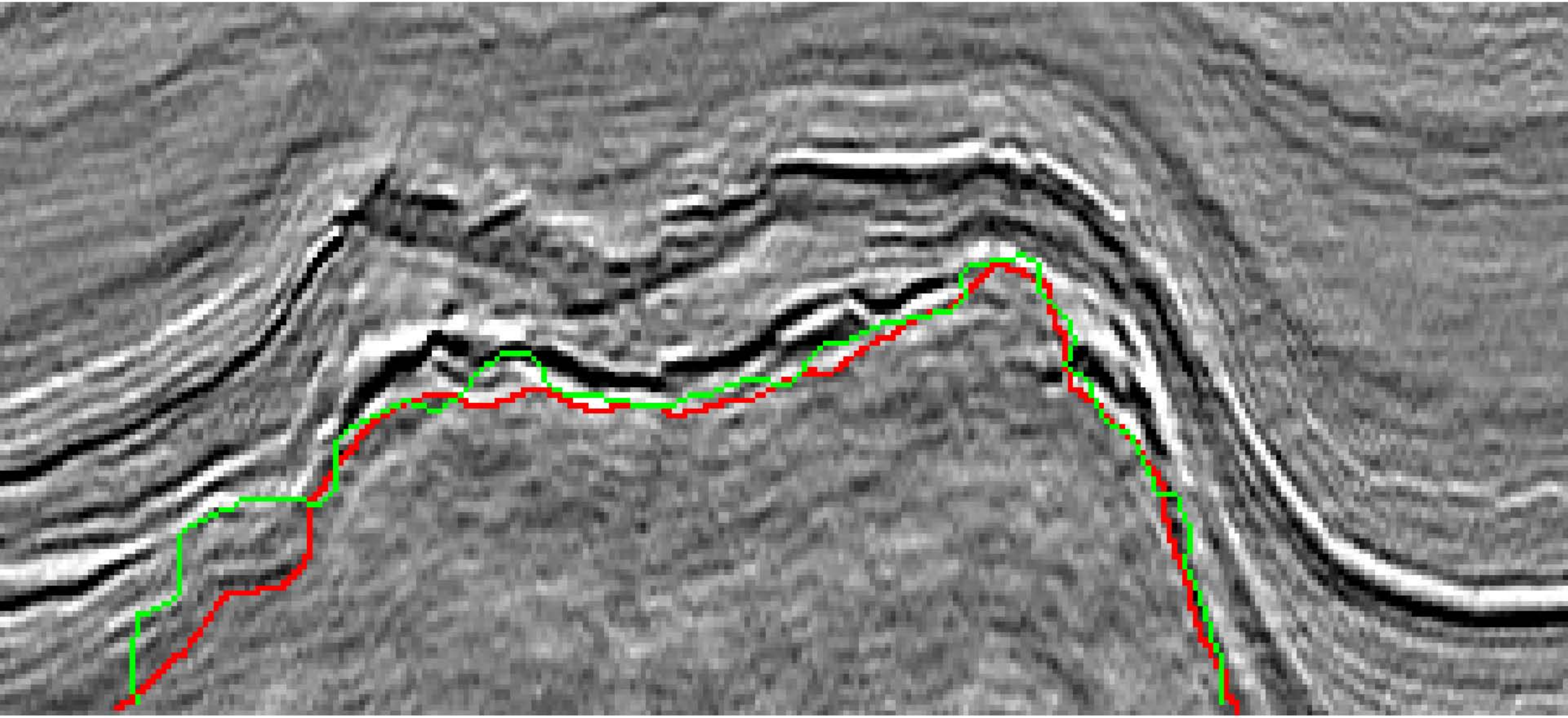}}
  \centerline{(c)}\medskip
\end{minipage}
\begin{minipage}[b]{0.48\linewidth}
  \centering
  \centerline{\includegraphics[height=3.3cm]{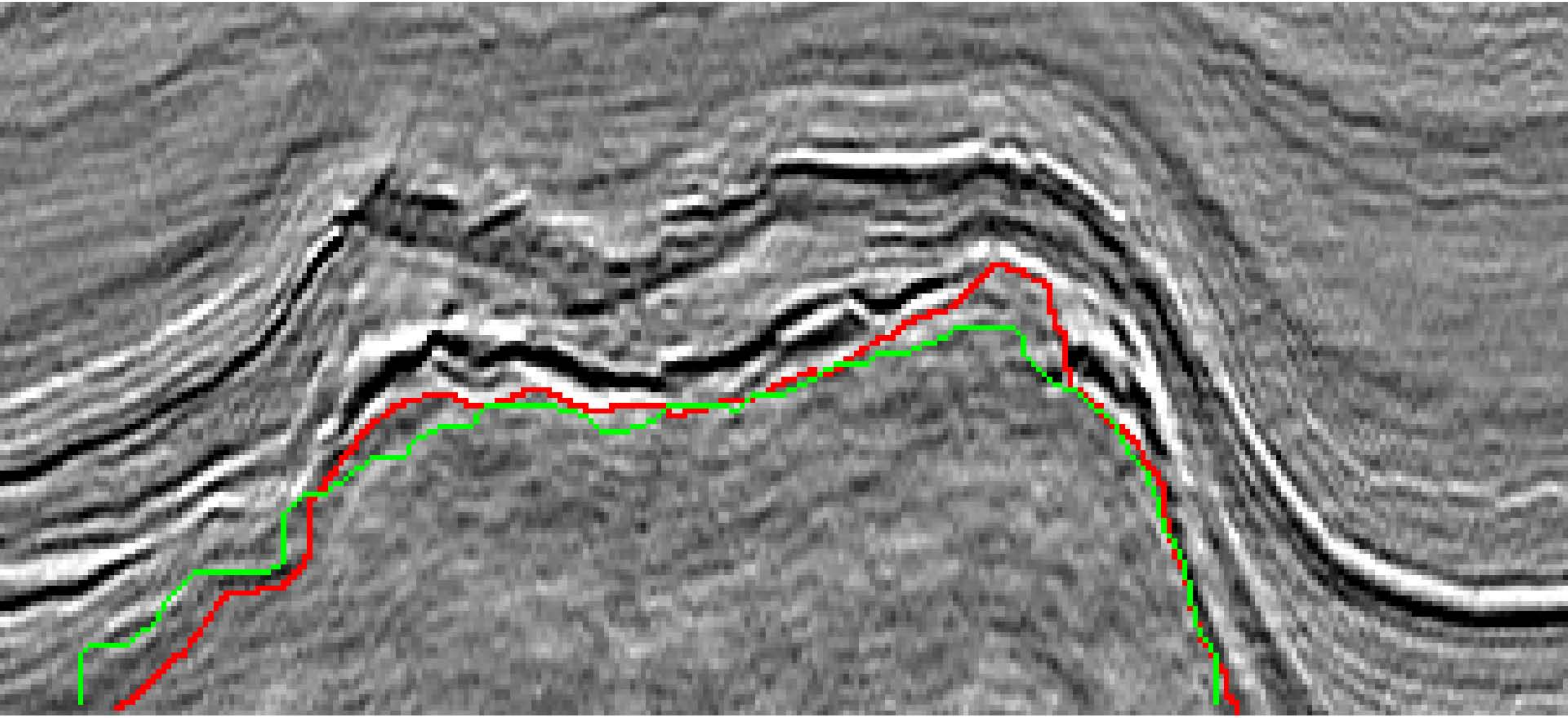}}
  \centerline{(d)}\medskip
\end{minipage}
\caption{(a) Tracked boundary of Inline \#392, (b) detected boundary of Inline \#392, (c) tracked boundary of Inline \#408, and (d) detected boundary of Inline \#408.}
\label{fig:tracking_onDetection}
\end{figure*}

In the previous sections, tracked boundaries are synthesized based on the boundary labeled by interpreters in the reference section. To further reduce human intervention in the tracking process, in the local seismic volume, we can synthesize the tracked boundaries of Inline \#389 to \#409 on the basis of the reference boundary in Inline \#400 labeled by the proposed detection method. Figure~\ref{fig:tracking_onDetection} compares the tracked and detected green boundaries with the red ground truth. In Figure~\ref{fig:tracking_onDetection}(a) and (b), the tracked boundary of Inline \#392 is more similar to the ground truth than the detected result. In contrast, in Figure~\ref{fig:tracking_onDetection}(c) and (d), since the bottom-left corner of the tracked boundary has great deviation to the ground truth, the detected boundary in Inline \#408 seems to be more accurate. However, the appearance of the great deviation is caused by the deviation in the reference boundary shown in Figure~\ref{fig:detection}(b) rather than our tracking method. To objectively compare the tracked and detected boundaries, we illustrate the corresponding SalSIM indices in Figure~\ref{fig:tracking_onDetection_cmpr}. Since we employ $\lambda_{\mathcal{C}}$ as an important weight that helps move tracked points to the boundary area, the tracked boundaries may have higher SalSIM indices than the detected ones, particularly in Inline \#395 to \#399. The mean of the SalSIM indices of the tracked boundaries is $0.9362$, which is almost the same as those of the detected boundaries, $0.9348$. It shows that the proposed tracking method is robust and efficient enough to delineate salt dome boundaries.


\subsubsection{Influence of Noise on Proposed Tracking Method}
\label{sssec:noiseadjusted}
Since in some cases the noise level in the reference section may be different from those of predicted sections, to eliminate the influence of noise on the tracked process, we propose the noise-adjusted tracking method, which selects PCs corresponding to the greatest SNR other than the greatest variation. In the local volume, we manually add gaussian noise with different variances to seismic sections. In the reference section Inline \#400, the added zero-mean gaussian noise has a variance of $0.01$. In contrast, the noise level increases to $0.05$ in predicted sections. We utilize the manually labeled reference boundary to synthesize tracked boundaries in Inline \#389 to \#409. In Figure~\ref{fig:noiseAdjusted}, We compare boundaries synthesized by the tensor-based tracking method with or without noise adjusted. With the increasing offset between the reference and predicted sections, the appearance of noise magnifies the error in the localization of tracked points. Since the noise-adjusted tracking method is robust to noise, the difference of two curves in Figure~\ref{fig:noiseAdjusted} grows with the increasing offset. The tracked boundaries obtained by the former one have a mean of SalSIM indices, $0.9468$, which is greater than that of the boundaries synthesized by the latter one, $0.9396$. It shows that the proposed noise-adjusted tracking method can alleviate the influence of noise and enhance the robustness of the tracking process.

\section{Conclusion}
In this paper, we propose noise-robust algorithms for detection and tracking of salt domes in post-migrated seismic volumes. By utilizing a texture-based gradient and a 2D DFT-based dissimilarity measure, we are able to accomplish noise-robust salt dome detection with both high accuracy and excellent efficiency. Then, by tracking a detected boundary from an initial seismic section through a 3D volume, we further improve the efficiency of the labeling process. The tensor-based noise-adjusted subspace learning approach proved very useful for the tracking of salt dome boundaries, especially when noise of varying levels is present. Combining together, the detection and tracking algorithms yield a practical salt dome labeling system that is effective, efficient, and robust.

\section{ACKNOWLEDGMENTS}
\label{sec:ack}
This work is supported by the Center for Energy and Geo Processing (CeGP) at Georgia Tech and King Fahd University of Petroleum and Minerals (KFUPM). 

\bibliography{main}
\bibliographystyle{seg}  

\clearpage
\plot*{diagram}{width=\columnwidth}
{An overall block diagram of the proposed salt-dome interpretation method.}

\plot{SaltDome}{width=0.6\columnwidth}
{The boundary of a salt dome in one seismic section is composed of the cap rock and side boundaries.}

\plot{GoT}{width=\columnwidth}
{The GoT at a given point represents the dissimilarity between two square neighborhood windows.}

\plot{GoT_xycomp}{width=0.65\columnwidth}
{The GoT is calculated along both the horizontal and vertical directions.}

\plot{neighb_wins}{width=0.6\columnwidth}
{Neighborhood windows of minimum size ($3\times 3$) and maximum size ($11\times 11$) shown around the boundary of a salt dome.}

\plot{sn_vectors}{width=0.7\columnwidth}
{Noise impact on the absolute value of the complex Fourier coefficients.}

\plot{unfolding}{width=\columnwidth}
{The unfolding of a third-order tensor along three modes.}

\plot{classification}{width=\columnwidth}
{The block diagram of the adaptive classification of texture tensors.}

\plot{localization}{width=\columnwidth}
{The block diagram of the identification of tracked points.}

\plot{localDist}{width=\columnwidth}
{An example of the salt dome boundary labeled by a computer-aided interpretation method and the ground truth labeled manually.}

\plot{Inline400}{width=\columnwidth}
{A seismic section (Inline \#400) of the local volume contains the cross-section of a salt dome.}
\clearpage

\plot{detection_cmpr}{width=\columnwidth}
{The SalSIM indices of salt dome boundaries detected from GoT, GLCM contrast, and gradient maps.}

\plot{crossline}{width=\columnwidth}
{A seismic section (Crossline \#834) of the crossline volume contains the cross-section of a salt dome.}

\plot{tracking_cmpr}{width=\columnwidth}
{The SalSIM indices of salt dome boundaries labeled by the proposed detection method and tracked boundaries based on detected boundary in Inline \#400.}

\plot{tracking_onDetection_cmpr}{width=\columnwidth}
{The SalSIM indices of tracked boundaries in predicted sections synthesized based on different tracking strategies.}

\plot{noiseAdjusted}{width=\columnwidth}
{The Comparison of tracked boundaries using tensor-based tracking method with and without noise adjustment.}

\end{document}